\documentclass[twocolumn]{aastex61}

\usepackage{natbib}
\usepackage{float}
\usepackage{appendix}

\begin{document}

%\title{Direct Imaging Constraints on the (Non-)Existence of Protoplanets Around LkCa 15 from SCExAO/CHARIS and Keck/NIRC2}
%\title{Evidence for the Non-Existence of Protoplanets Around LkCa 15\\  from SCExAO/CHARIS and Keck/NIRC2}
\title{Laboratory Demonstration of Spatial Linear Dark Field Control For Imaging Extrasolar Planets in Reflected Light}
 % \\LkCa 15 \lowercase{bcd} are Likely (Contaminated By) Inner Disk Signals}
\correspondingauthor{Thayne Currie}
\email{currie@naoj.org,thayne.m.currie@nasa.gov}
\author{Thayne Currie}
\affiliation{NASA-Ames Research Center, Moffett Blvd., Moffett Field, CA, USA}
\affiliation{Subaru Telescope, National Astronomical Observatory of Japan, 
650 North A`oh$\bar{o}$k$\bar{u}$ Place, Hilo, HI  96720, USA}
\affiliation{Eureka Scientific, 2452 Delmer Street Suite 100, Oakland, CA, USA}
\author{Eugene Pluzhnik}
\affiliation{NASA-Ames Research Center, Moffett Blvd., Moffett Field, CA, USA}
\affiliation{Bay Area Environmental Institute, P.O. Box 25, Moffett Field, CA 94035-0001}
\author{Olivier Guyon}
\affiliation{Subaru Telescope, National Astronomical Observatory of Japan, 
650 North A`oh$\bar{o}$k$\bar{u}$ Place, Hilo, HI  96720, USA}
\affil{Steward Observatory, University of Arizona, Tucson, AZ 85721, USA}
\affil{College of Optical Sciences, University of Arizona, Tucson, AZ 85721, USA}
\affil{Astrobiology Center of NINS, 2-21-1, Osawa, Mitaka, Tokyo, 181-8588, Japan}
\author{Ruslan Belikov}
\affiliation{NASA-Ames Research Center, Moffett Blvd., Moffett Field, CA, USA}
\author{Kelsey Miller}
\affiliation{Leiden Observatory, Leiden University, P.O. Box 9513, 2300 RA Leiden, The Netherlands}
\author{Steven Bos}
\affiliation{Leiden Observatory, Leiden University, P.O. Box 9513, 2300 RA Leiden, The Netherlands}
%\author{Frans Snik}
%\affiliation{Leiden Observatory, Leiden University, Leiden, The Netherlands}
\author{Jared Males}
\affiliation{Steward Observatory, University of Arizona, Tucson, AZ 85721, USA}
\author{Dan Sirbu}
\affiliation{NASA-Ames Research Center, Moffett Blvd., Moffett Field, CA, USA}
\author{Charlotte Bond}
\affiliation{Institute for Astronomy, University of Hawaii,
640 North A`oh$\bar{o}$k$\bar{u}$ Place, Hilo, HI 96720, USA}
\author{Richard Frazin}
\affiliation{Department of Climate and Space Sciences and Engineering, University of Michigan-Ann Arbor, MI 48109-2143, USA}
\author{Tyler Groff}
\affiliation{NASA-Goddard Space Flight Center, Greenbelt, MD, USA}
\author{Brian Kern}
\affiliation{Jet Propulsion Laboratory, California Institute of Technology, 4800 Oak Grove Drive, Pasadena, California 91016, USA}
\author{Julien Lozi}
\affiliation{Subaru Telescope, National Astronomical Observatory of Japan, 
650 North A`oh$\bar{o}$k$\bar{u}$ Place, Hilo, HI  96720, USA}
\author{Benjamin A. Mazin}
\affiliation{Department of Physics, University of California, Santa Barbara, CA 93106, USA}
\author{Bijan Nemati}
\affiliation{Jet Propulsion Laboratory, California Institute of Technology, 4800 Oak Grove Drive, Pasadena, California 91016, USA}
\author{Barnaby Norris}
\affiliation{Sydney Institute for Astronomy, School of Physics, Physics Road, University of Sydney, NSW 2006, Australia}
\author{Hari Subedi}
\affiliation{Department of Mechanical and Aerospace Engineering, Princeton University, Princeton, NJ, USA}
\author{Scott Will}
\affiliation{NASA-Goddard Space Flight Center, Greenbelt, MD, USA}
\affiliation{The Institute of Optics, University of Rochester, Rochester, NY, USA} 

\begin{abstract}
Imaging planets in reflected light, a key focus of future NASA missions and ELTs, requires advanced wavefront control to maintain a deep, temporally correlated null of stellar halo -- i.e. a dark hole -- at just several diffraction beam widths.   Using the Ames Coronagraph Experiment testbed, we present the first laboratory tests of Spatial Linear Dark Field Control (LDFC) approaching raw contrasts ($\sim$ 5$\times$10$^{-7}$) and separations (1.5--5.2 $\lambda$/D) needed to image jovian planets around Sun-like stars with space-borne coronagraphs like WFIRST-CGI and image exo-Earths around low-mass stars with future ground-based 30m class telescopes.   %LDFC uses the response to perturbations in uncorrected, 'bright field' regions to maintain a dark hole without continuous DM probing.
In four separate experiments and for a range of different perturbations, LDFC largely restores (to within a factor of 1.2--1.7) and maintains a dark hole whose contrast is degraded by phase errors by an order of magnitude.   Our implementation of classical speckle nulling requires
%,  LDFC typically converges to within 
%a factor of 1.2--1.7 of the original dark hole intensity in 
a factor of 2--5 more iterations and 20--50 DM commands to reach contrasts obtained by spatial LDFC.   Our results provide a promising path forward to maintaining dark holes without relying on DM probing and in the low-flux regime, which may improve the duty cycle of high-contrast imaging instruments, increase the temporal correlation of speckles, and thus enhance our ability to image true solar system analogues in the next two decades. 
\end{abstract}
\section{Introduction}
Over the past decade, ground-based telescopes using facility adaptive optics (AO) systems and now dedicated \textit{extreme} AO systems have provided the first direct images of self-luminous, (super-)jovian mass planets orbiting young stars \citep[e.g.][]{Marois2008,Lagrange2010,Rameau2013,Currie2014,Currie2015,Macintosh2015, Chauvin2017,Keppler2018}.  Follow-up multi-wavelength photometry and spectroscopy \citep{Currie2011,Barman2015,Rajan2017} have yielded the first constraints on their atmospheric properties, such as clouds, chemistry and surface gravity.  The soon-to-be launched \textit{James Webb Space Telescope} may provide the first direct images of self-luminous (super-)jovian exoplanets around intermediate-aged stars and will prove a unique probe of atmospheric chemistry and the properties of dust entrained in exoplanets' clouds \citep[e.g.][]{Beichman2010}.

%However, directly detecting and characterizing the spectrum of a habitable zone Earth-like planet around a Sun-like star with a future space mission requires suppression of noisy, scattered halo of starlight better by a factor of 10$^{10}$ (Figure \ref{fig:contrast}).  Planet-to-star contrasts required to image a  habitable-zone Earth around an M star are a factor of 100 milder ($\sim$ 10$^{-7}$--10$^{-8}$) but require detection at $\sim$ 20--100 milliarcseconds, typical $\approx$ 1--5$\lambda$/D at near-infrared wavelengths capable of revealing evidence for biomarkers (e.g. the 1.27 $\mu$m oxygen line\citep{LopezMorales2019}).  

%\textbf{\textit{STUFF NOT DONE YET: 
%\begin{itemize}
%\item show sequence of subtractions image - simulated point source, image - simulated - extended source (Fig 3, right)
%\item Finish photometry for HD 100546 c
%\item image showing predicted position of HD 100546 c/epoch or simulated pt source/extended source subtraction (Fig 4. right)
%\item Get prediction for HD 100546 c in the near-future
%\end{itemize}}}

Imaging exoplanets in reflected light from future space missions or ground-based extreme AO systems requires new advances in wavefront control (WFC) and coronagraphy \citep[e.g.][]{Guyon2018c,Crill2019}.  High-contrast imaging testbeds utilizing focal plane WFC techniques like speckle nulling \citep{Malbet1995,Borde2006} and electric field conjugation (EFC; \citep{Give'on2007}) and advanced coronagraphy can generate deep dark holes (DH) around a star at the 10$^{-8}$ level in air and 10$^{-9}$ or lower in vacuum \citep[e.g.][]{Trauger2011,Belikov2011,Belikov2012,Cady2016}.  On ground-based telescopes, wavefront sensing and control advances (e.g. Zernike phase sensing and predictive control) have shown promise on new, state-of-the-art extreme AO systems like SCExAO and could yield orders of magnitude gain in raw contrast \citep{Ndiaye2013,Ndiaye2016,MalesGuyon2018,Males2018,Currie2019a}.  

Achieved null depths in monochromatic light and narrow bandpasses ($\lesssim$ 10$^{-8}$--10$^{-9}$) are, \textit{if sustained}, sufficient to image reflected-light jovian planets orbiting at 1--5 au from space telescopes, even around obscured apertures like WFIRST-CGI \citep{Seo2018,Shi2018}.  On the ground, upcoming \textit{extremely large telescopes} (ELTs) delivering sustained contrasts of $\sim$ 10$^{-6}$ and temporally-correlated residual speckles \textit{could} enable reflected-light imaging of numerous jovian planets, even perhaps Earth-like planets around the nearest low-mass stars.
 \citep[e.g.][]{Guyon2018a,LopezMorales2019}.

Sustaining deep contrasts within a DH necessary to image planets in reflected light imposes significant demands on wavefront sensing, as the residual stellar halo must be measured with extreme precision.   Precision sensing is particularly difficult  when the DH itself is used for focal-plane wavefront control (FPWFC) and is already photon starved, as can be the case for standard methods like EFC and speckle nulling.  Furthermore, by modulating the deformable mirror (DM) to determine and update an estimate of the electric field, FPWFC methods like EFC can perturbe science exposures and thus limit an observation's duty cycle.   Instead of using the science target for FPWFC itself, another strategy (for WFIRST CGI) is to first dig a DH around a far brighter reference star within 15-20$^{o}$ of a science target and then apply the high-order DM correction to the science target \citep{Bailey2018}.  However, both the average contrast of the DH and its temporal correlation with respect to its initial state can and likely will degrade due to any number of dynamic aberrations.   Slewing back to the reference star to rebuild the DH, as is currently baselined for WFIRST-CGI, substantially increases an observation's duty cycle.  Advanced post-processing methods can yield substantial contrast gains (a factor of $\sim$ 100) \citep[e.g.][]{Soummer2011,Currie2012}.   But the brightening of the DH and its decorrelation over time degrades the effectiveness of these post-processing methods to remove residual starlight impeding planet detection.
 
    \begin{figure} [ht]
   \begin{center}
  % \begin{tabular}{c} %% tabular useful for creating an array of images 
  %\centering
   \includegraphics[width=0.5\textwidth,clip]{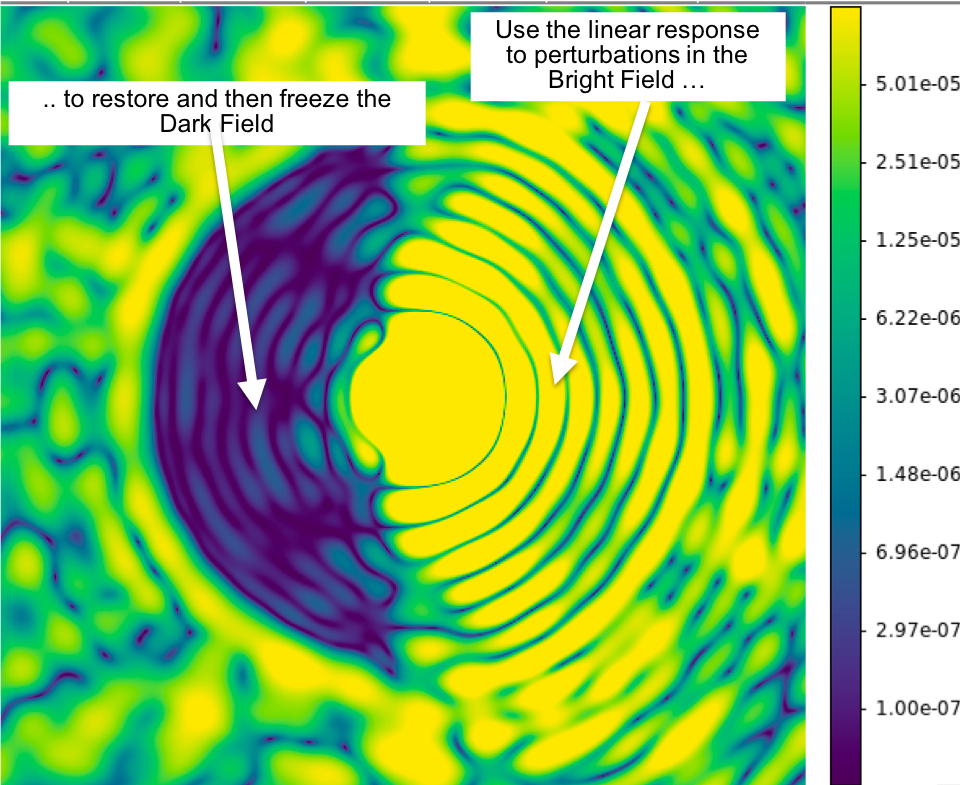}
  % \end{tabular}
   \end{center}
   \vspace{-0.2in}
   \caption
%>>>> use \label inside caption to get Fig. number with \ref{}
   { \label{fig:ldfcschematic} 
   Schematic of Spatial Linear Field Dark Control obtained from simulated data for the Ames Coronagraph Experiment testbed.   Bright, uncorrected regions with a contrast with respect to the peak intensity of $\sim$ 10$^{-4}$ are used to stabilize a dark hole with a contrast of $\sim$ 10$^{-7}$--10$^{-8}$.}
   \end{figure}

Linear Dark Field Control (LDFC) is a promising wavefront control method which could maintain a static, deep DH without deformable mirror probing after the DH's creation from FPWFC methods \citep{Miller2017}.   LDFC utilizes the linear response of the uncorrected but photon-rich region in the focal plane (the ``bright field" or BF) to wavefront perturbations that affect both the BF and the photon-starved DF\footnote{Throughout, ``corrected" means "corrected for aberrations using focal-plane wavefront sensing and control techniques"}.   Because LDFC does not require modulating the signal within the DH, it needs  only a single focal plane image to restore the electric field to its initial state.
%correct for the change in the electric field from the initial state. 

LDFC can be implemented in at least two ways. ``Spatial" LDFC in a single band image, where a DH is created on one side of the image and stabilized by the BF on the opposite side \citep{Miller2017} as shown in Figure \ref{fig:ldfcschematic}.  "Spectral" LDFC where the BF draws from pixels in out-of-band image slices at wavelengths bracketing the bandpass within which the DH is created \citep{Guyon2017}.    

\citet{Miller2017} and \citet{Guyon2017} presented numerical simulations showing that an LDFC control loop should be able to hold static a DH at a 10$^{-7}$--10$^{-8}$ contrast level similar to that initially created using FPWFC methods.  \citet{Miller2019} presented numerical simulations and early laboratory tests demonstrating LDFC coupled with the vector apodized phase plate coronagraph at  10$^{-3}$ contrast between 4 and 11 $\lambda$/D; \citet{Currie2019b} presented preliminary results from the Ames Coronagraph Experiment testbed showing that spatial LDFC may be successful at partially-restoring a DH at 10$^{-5}$ contrast in some cases.  While encouraging, the tests were compromised by bright, static uncorrectable regions left on the focal plane images due to system internal reflection and non-ideal regularization of the control matrix used to map between changes in the focal plane and changes in the DM shape.   Deeper contrasts (10$^{-6}$--10$^{-7}$) are needed to test Spatial LDFC in regimes important for imaging exoplanets in reflected light from upcoming ground-based telescopes and space missions and better determine the limitations of LDFC.
%to possible null space between the response of the dark and bright fields.      

%Linear Dark Field Control\citep{Miller2017} (LDFC) is a promising wavefront control method which could maintain a static, deep DH that is first generated from FPWFC methods like EFC.   It utilizes the linear response of the uncorrected region in the focal plane which has far larger signal (the ``bright field" or BF) to wavefront perturbations that affect both the BF and the DF.   LDFC does not require modulating the signal within the DH and therefore requires only a single focal plane image to correct for the change in the electric field from the initial EFC-estimated state.    LDFC can be implemented in at least two ways: 1) ``Spatial" LDFC in a single band image, where a DH is created on one side of the image and stabilized by the BF on the opposite side\citep{Miller2017} as shown in Figure \ref{fig:ldfcschematic} or "Spectral" LDFC where the BF draws from pixels in image slices at wavelengths bracketing the bandpass within which the DH is created\citep{Guyon2017}.   Numerical simulations show that an LDFC control loop is able to hold static a DH at a contrast level comparable to that initially created using EFC, motivating laboratory tests to validate its efficacy at contrast levels needed for ground-based imaging of rocky planets with ELTs ($\sim$ 10$^{-5}$--10$^{-6}$ raw contrast) and later for WFIRST-CGI or NASA flagship missions like HabEx or LUVOIR ($\sim$ 10$^{-8}$--10$^{-10}$).  

In this work, we present the first laboratory demonstration of Spatial Linear Dark Field Control at contrasts relevant for future imaging of exoplanets in reflected light, using the Ames Coronagraph Experiment (ACE) testbed \citep{Belikov2009}.   After briefly reviewing the premise of LDFC (\S 2), we describe our experimental setup for testing Spatial LDFC at ACE (\S 3) at contrasts relevant for imaging reflected-light planets but shallow enough that phase errors dominate the wavefront error budget \citep{ShaklanGreen2006,Pueyo2007}.   \S 4 describes our results, where the LDFC control loop is used to largely restore dark holes that are corrupted by a range of different perturbations and its performance is benchmarked against our implementation of a classical speckle nulling loop.   The discussion (\S 5) details plans to further benchmark LDFC, testing its performance in regimes similar to those that will be faced with WFIRST-CGI, and sketches ways to implement a version of LDFC with WFIRST-CGI and future ground-based telescopes.
%comparable to those 
%In this work, we describe the first steps to validate and mature Spatial LDFC (hereafter "LDFC") in a laboratory setting using the Ames Coronagraph Experiment (ACE) testbed, at contrast levels relevant for direct imaging of mature exoplanets in reflected light with upcomig telescopes like the \textit{European Extremely Large Telescope} (E-ELT), \textit{Giant Magellan Telescope} (GMT), and \textit{Thirty Meter Telescope}.    

\section{Linear Dark Field Control Background}
%\subsection{LDFC Theory}
Linear Dark Field Control theory was first described in \citet{Miller2017}.   The premise of LDFC is that perturbations in the pupil plane induce a response in the electric field in both the corrected, deep-contrast regions of the focal plane (the ``dark field"; DF) and uncorrected, shallow-contrast regions (the ``bright field"; BF).  For spatial LDFC, this premise is generally true if phase errors dominate the wavefront error budget, as they produce aberrations in both the DF and BF regions (e.g. a sine wave perturbation on a DM producing a pair of speckles)\footnote{\S 5 discusses the applicability and possible null space of LDFC in an ultra-deep contrast regime where amplitue errors become important.}.   This perturbation is small compared to the bright field intensity; the response to changes in the bright field is then linear, not quadratic.  Given an influence function (e.g. a response matrix) that describes the mapping between the DM shape and changes in the bright field (with respect to its unperturbed state), one can find a unique solution to the change in DM shape that restores both the unperturbed bright field and dark field.  

   Briefly, the electrical field in focal plane at a given time $t$ can be described as the sum of the incident electric field $E_{\rm o}$ established by focal-plane wavefront sensing techniques and a small change in complex amplitude due to a small wavefront error in a conjugate plane, $E_{\rm 1}$, that corrupts the DH and could be corrected by the DM:
   %induced by the deformable mirror, $E_{\rm 1}$:
\begin{equation}
\label{eq:efield}
%\abs{E} \approx 1
E_{\rm t} \approx E_{\rm o} + E_{\rm 1}.
%2 a = \frac{(b + 1)}{3c} \, ,
\end{equation}

The intensity in the focal plane, $I_{\rm t}$ = $|E_{\rm t}|^{2}$, is then comprised of three terms.  These are the intensity due to the initial electric field, the intensity due to changes in the electric field due to the wavefront error, $E_{\rm 1}$, and the inner product between the two electric field components:
\begin{equation}
\label{eq:i3terms}
%\abs{E} \approx 1
|I_{\rm t}| \approx |E_{\rm o}|^{2} + |E_{\rm 1}|^{2} + 2<E_{\rm o},E_{\rm 1}>.
%2 a = \frac{(b + 1)}{3c} \, ,
\end{equation}

Within the dark field at time $t$, $|E_{\rm 1}|^{2}$ dominates, as the initial electric field component $E_{\rm o}$ is small.   However, $E_{\rm o}$ is primarily responsible for the intensity distribution in the bright, uncorrected region: $|E_{\rm o}|^{2}$ $\gg$ $|E_{\rm 1}|^{2}$ and 2$<E_{\rm o},E_{\rm 1}>$ $\gg$ $|E_{\rm 1}|^{2}$.   Therefore, the change in the focal plane intensity $I$ between time 0 and $t$ in the BF is a linear function of the change in complex amplitude induced by changes in the electric field due to the wavefront error: $\Delta$$I$ = $I_{\rm t}$ - $I_{\rm o}$  $\approx$ $2<E_{\rm o},E_{\rm 1}>$.

  \begin{deluxetable*}{llllll}
%\setlength{\tabcolsep}{0pt}
%\tablecaption{Preliminary SCExAO/CHARIS Astrometric Calibration}
\tablecaption{Experiment Log\label{tab:experiment}}
\tablewidth{0pt}
%\tablenum{2}
%\tablehead{\colhead{Name} & \colhead{$\chi^{2}_{\nu}$} & \colhead{$\chi^{2}_{\nu, H+K}$}&\colhead{SpT} & \colhead{H$_{\rm cont.}$ index} & \colhead{H$_{2}$K index} & \colhead{Mov. Group} & \colhead{Age} & \colhead{log(L/L$_{\odot}$)}&\colhead{Mass (M$_{\rm J}$)}  & \colhead{References} \\
\tablehead{\colhead{Experiment Number} & \colhead{Date} & \colhead{Dark Field Size} & \colhead{Bright Field Size} & \colhead{Starting Dark Hole Contrast} &  \colhead{Aberrations} }
\tiny
\centering
\startdata
1&2019-09-19 & 1.6--5.1 $\lambda$/D & 1.4--5.1 $\lambda$/D & 6.50$\times$10$^{-7}$ & single speckle \\
% & & & &15&single actuator\\
 %& & & &15&linear comb. sine wave\\
 %& & & &15&Kolmogorov phase screen\\
 2& 2019-12-23 & 1.6--5.1 $\lambda$/D& 1.6--5.1 $\lambda$/D & 5.97$\times$10$^{-7}$ & two speckles\\
% & & & & sine wave\\
 %& & & & linear comb. sine wave\\
 %& & & & Kolmogorov phase screen\\
 3 & 2019-01-12 & 1.65--5.2 $\lambda$/D& 1.55--5.2 $\lambda$/D & 6.85$\times$10$^{-7}$ & low spatial frequency\\
 4& 2019-01-25 & 1.65--5.2 $\lambda$/D& 1.65--5.2 $\lambda$/D &  4.97$\times$10$^{-7}$ & complex/three speckles\\
 %& & & & sine wave\\
 %& & & & linear comb. sine wave\\
 %& & & & Kolmogorov phase screen\\
\enddata
\vspace{-0.05in}
\tablecomments{Starting Dark Hole Contrast refers to the average intensity within the DH with respect to the peak signal from the laser source.   The average intensity over the (smaller) scoring regions for Experiments \#2-4 is comparable.}
%\label{tab:experiment}
\vspace{-0.2in}
\end{deluxetable*}

By 1) measuring changes in the bright field intensity between time $t_{\rm o}$ when the DH is first established and time $t$ where it is corrupted and 2) constructing an influence function mapping between DM shape and focal plane intensity, we can then determine the set of DM actuator offsets that restore both the initial bright field and initial dark field corrupted by phase errors.        

We adopt a system response matrix, $RM$, with dimensions of $n$ bright field pixels by $m$ actuators.   The $RM$ links together changes in DM shape $\Delta u_{\rm t}$ to changes in the bright field intensity distribution: $\Delta$$I_{\rm DM, t}$ = $RM$$\Delta u_{\rm t}$.   Actuator offsets $\Delta u_{\rm t}$ required to drive the dark field back to its original state at time $t$ are then equal to the pseudo-inverse of $RM$ (i.e. the ``control matrix", $CM$) multiplied by the change in the bright field, $\Delta$$I_{\rm BF}$:
\begin{equation}
    \Delta u_{\rm t} = -(RM^{T}RM)^{-1}RM^{T}\Delta I_{\rm t, BF}.
\end{equation}

LDFC has two potential key advantages over DM probing methods like EFC and speckle nulling, which use measurements of the DH directly for sensing and control.   First, the signal within the photon-rich (uncorrected) bright field is larger than the (corrected) dark field and is not impacted by camera readout noise.   Thus,  for extremely deep-contrast DHs where the residual DH signal is photon starved, LDFC provides a higher signal-to-noise measurement of the DM shape needed to maintain/freeze the DH initial state.   
%In Appendix A, we verify that the differential signal within the bright field is in fact significantly larger than that within the dark field below DH contrasts of 10$^{-6}$.  

Second, LDFC is a differential wavefront control technique.  Once the $CM$ for LDFC is determined, LDFC requires a single focal plane measurement to determine the change in DM shape that will restore the DH.   In comparison, methods like EFC and speckle nulling rely on DM probing to determine the change in DM shape that will eliminate perturbations within the DH.   Probing requires introducing perturbations in the pupil plane to determine the phase of speckles: e.g., for our implementation of classical speckle nulling, 4-6 probes must be introduced to solve for the speckle phase.   Thus, even if the same number of iterations allow LDFC and methods like EFC/speckle nulling to restore the DH, the duty cycle for LDFC could be significantly shorter.   
%Even if DM probing methods restore an equivalent DH average intensity, they do not necessary restore the same spatial structure of residual speckles.   

%For a small pure phase abberation introduced by the DM, equation (3) can be written as 
%\begin{equation}
%    \Delta I_{\rm DM} = RM\times u,
%\end{equation}
%    where RM is the system response (Jacoby) matrix with dimensions of n image plane pixels and m DM actuators, and $u$ offsets, required to change the BF focal plane intensities by an amount $\Delta I_{\rm DM}$.   Assuming that bright field rintensity recovery to its initial state $I_{\rm o}$ simultaneously recover the dark field, the actuators offsets required to drive the dark field back to its original state can be found as the solution of a linear system of equations
  %  \begin{equation}
 %       \Delta I_{\rm DM} = -\Delta I_{\rm t} = RM u_{t},
 %   \end{equation}
%can be written as $u_{\rm t}$ = CM$\Delta$ $I_{\rm t}$, where CM=-($RM^{T}$$RM$)$^{-1}$$RM^{T}$ is the control matrix is equal to the pseudoinverse of CM.
% \newpage

\section{Linear Dark Field Control Experiments}
We conducted tests of LDFC using the Ames Coronagraph Experiment (ACE) laboratory at NASA-Ames Research Center in four separate experiments between September 2019 and January 2020 (Table \ref{tab:experiment}) at contrast levels shallow enough that phase errors are expected to dominate but deep enough to be relevant for future ground and space high-contrast imaging.   

Our specific experiment milestone was as follows:
\begin{itemize}
    \item   For a DH at a starting raw contrast of $\approx$ 10$^{-6}$ that is degraded by at least a factor of 10 by injected phase perturbations over at least a region with an area of $\sim$ 10 ($\lambda$/D$)^{2}$, demonstrate that 1) spatial LDFC can yield at least a 10x gain in DH contrast and thus largely restore the DH and 2) hold this gain for over 100 iterations.   
    \item Achieve at least three successful demonstrations of achievements 1) and 2).
    \end{itemize}
%and 3) repeat this experiment at least two additional times (a total of at least 3 demonstrations).   

\subsection{Laboratory Setup}
  The testbed uses a laser centered on 635nm as a monochromatic light source.  To limit file size and improve the speed of the wavefront control loop, we read out focal-plane images in 700x700 or 500x500 subarrays.   The 1 $\lambda$/D full-width-at-half-maximum point-spread function (PSF) size measured $\sim$ 32 pixels.  

For each experiment, satellite speckles were used to determine the conversion factor between counts and contrast with respect to the peak of an unocculted PSF.  We used the PIAA coronagraph to suppress scattered starlight \citep{Guyon2003,Guyon2010} and a circular occulting spot of $\sim$ 1 $\lambda$/D radius to yield a full 360 degree spatial coverage.      To achieve an initial flat wavefront at the pupil plane, we use an implementation of the Gerchberg-Saxton method, which solves for the flat DM shape using a sequence of random pupil plane phase probes \citep{Pluzhnik2017}.  
%The D-shaped focal plane occulter normally blocking all of one side of the wavefront sensor image was removed, making visible the full 360 degree field needed for LDFC.
   
%To create a one sided, C-shaped dark hole extending between 1.2 $\lambda$/D and 5.5 $\lambda$/D
We used a classical speckle nulling control loop as implemented in previous ACE testbed experiments (e.g. \citep{Belikov2012}) to  correct for up to 81 speckles at a time.  For each iteration of speckle nulling, we issue on average 10 DM commands: 7 to determine the phase of the speckles and 3 to determine amplitude.   

The speckle nulling loop created a one sided, C-shaped DH extending from an inner working angle of 1.5--1.6 $\lambda$/D  to an outer working angle of 5.1--5.2 $\lambda$/D.   The average contrast within the DH measured between 4.97 $\times$10$^{-7}$ and 6.85$\times$10$^{-7}$ depending on the experiment.   In units of contrast, the approximate read-noise level of the detector was $\sim$ 5$\times$10$^{-7}$ for the September and December experiments and a factor of 2 lower for the January experiments due to a factor of $\approx$ 10 longer exposures for the latter.   Assuming a reasonable gain from post-processing (e.g. 30-50x), these raw contrasts are similar to the performance needed to detect jovian planets at $\sim$ 1 au in reflected light around nearby stars.
%around those expected for the half-dozen or so jovian planets at 0.5--2 au 

   \begin{figure} [ht]
   %\begin{center}
  % \begin{tabular}{c} %% tabular useful for creating an array of images 
  %\centering
   \includegraphics[width=0.5\textwidth,trim = 15mm 70mm 6mm 70mm,clip]{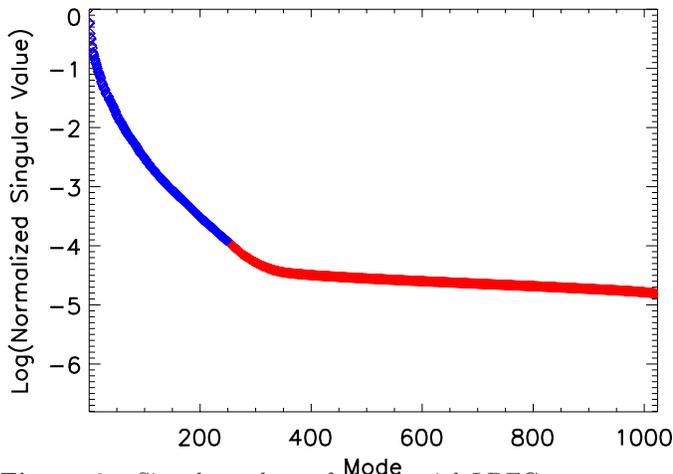}
  % \end{tabular}
  % \end{center}
   \vspace{-0.3in}
   \caption{Singular values of the spatial LDFC response matrix, $RM$, for Experiment $\#4$ (25 January 2020).   Out of 1024 total modes, 250 modes (blue) were retained in the control matrix calculation, while higher modes (red) were discarded. 
   }
   \label{fig:evals}
   \end{figure} 
\subsection{Linear Dark Field Control Matrix Setup and Closed-Loop Implementation}

%Due to our removal of the occulter, internally reflected light partially contaminates three isolated regions of the dark hole.   The average dark hole intensity contrast with respect to the PSF core is roughly $\sim$ 10$^{-5}$: a contrast below that currently achieved at 1--5 $\lambda$/D with extreme AO systems on 8-10m telescopes but approaching that required for imaging mature planets in reflected light around M stars with ELTs.   At 1.2--4.5 $\lambda/D$, the average intensity in the bright, uncorrected region ranges between 10$^{-3}$ and 10$^{-4}$.

To calculate the Spatial LDFC response matrix ($RM$), we 
%followed methods for collecting an $RM$ with the CACAO software \citep{Guyon2018b}, 
perturbed each of the $m$ actuators by a series of small amplitude pokes, 1 and 2,
%\footnote{Separate tests verified that our poke amplitudes kept the BF within the linear response regime.} 
which are performed sequentially and have opposite signs (positive and negative). We then recorded the intensity $I$ over $n$ BF pixels.   Each of the pokes have a fixed amplitude of $ampl_{\rm poke}$.  We combine results from two separate patterns -- $a$ and $b$ -- which differ by the order in which the positive/negative pokes are applied (i.e. $a$ = +- +-, $b$= +- -+):
\begin{equation}
    RM(n,m) = 0.5*[(I_{\rm a_{\rm 1}}-I_{\rm a_{\rm 2}})+(I_{\rm b_{\rm 1}}-I_{\rm b_{\rm 2}})]/(2*ampl_{\rm poke})
\end{equation}

The control matrix ($CM$) in a closed-loop implementation of LDFC is the pseudo-inverse of $RM$: 
\begin{equation}
    CM = (RM^{T}RM)^{-1}RM^{T}
\end{equation}.    

 \begin{figure} [ht]
  % \begin{center}
  % \begin{tabular}{c} %% tabular useful for creating an array of images 
  %\centering
   \includegraphics[width=0.499\textwidth,clip]{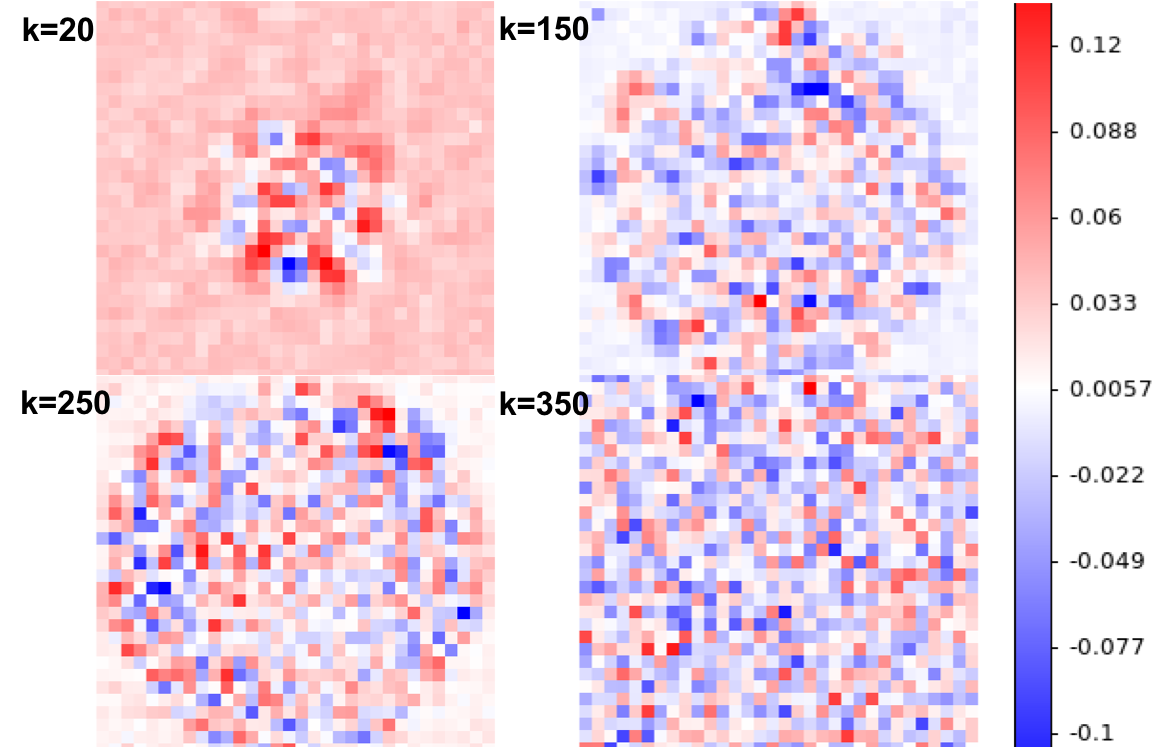}
  % \end{tabular}
  % \end{center}
   \vspace{-0.1in}
   \caption
%>>>> use \label inside caption to get Fig. number with \ref{}
   {DM modes from the Control Matrix calculation for Experiment $\#4$ (25 January 2020).  For modes lower than $k$ $\sim$ 250, the response is confined to within a circular region that roughly match the coronagraph pupil.   At modes higher than $k$ $\sim$ 250, the response varies at the pixel-to-pixel level. 
   %Schematic of Spatial Linear Field Dark Control obtained from simulated data for the Ames Coronagraph Experiment testbed.   Bright, uncorrected regions with a contrast with respect to the peak intensity of $\sim$ 10$^{-4}$ are used to stabilize a dark hole with a contrast of $\sim$ 10$^{-7}$--10$^{-8}$.
   }
   \label{fig:modalresponse}
   \end{figure} 
Since (5) is usually ill-conditioned, we apply a truncated SVD regularization when computing it.  Specifically, To compute $CM$, we decompose ($RM^{T}$$RM$)$^{-1}$ into a matrix of eigenvectors $V$ and a matrix of eigenvalues $\Lambda$,  truncating $\Lambda$ at mode $k_{lim}$ before inverting to yield the $CM$: 
$CM$ = ($V$${\Lambda}^{-1}$$V^{T}$)$_{\rm k<k_{lim}}$$RM^{T}$.

%$CM$ = ($V$${\Lambda}^{-1}$$V^{T}$)$RM^{T}$
%$CM$ = ($V$${\Lambda}^{-1}_{\rm k<k_{lim}}$$V^{T}$) $RM^{T}$. 
%$CM$ = ($V$$\Lambda}^{-1}_{\rm k<k_{lim}$$V^{T}$) $RM^{T}$.  

The normalized singular values of the $RM$ covariance decline to $\sim$ 10$^{-3}$ by $k$ = 200 and flatten to 10$^{-4}$ between $k$ = 250 and $k$ = 1024 (where the $RM$ covariance would be at full-rank) (Figure \ref{fig:evals}).   Inspection of the modal responses showed that signal at $k$ $>$ 300 was dominated by very high frequency pixel-to-pixel variations; at $k$ = 250, the response was still clearly dominated by spatially correlated signal (Figure \ref{fig:modalresponse}).   Thus, we set a modal cutoff to the $CM$ at $k$ = 250.  
%We set the modal cutoff to an intermediate value consistent with the highest mode where the imprint of a circular region of the DM (excluding the corners) is still visible 
%(Figure \ref{fig:svdlim}b).
  
  Our closed-loop implementation of LDFC multiplies the DM offset shape in the $i$-th iteration $\Delta u_{\rm t,i}$ by a gain $g$ and adds this value to the current DM shape: 
  %\begin{equation}
  %    DM_{\rm i} = DM_{\rm i-1}+u_{\rm t,i}\times~g
  %\end{equation}
  $DM_{\rm i} = DM_{\rm i-1}+\Delta~u_{\rm t,i}\times$~g. We tested a range of gain values.  For simplicity, we settled on $g$=0.25 for all experiments, which provided a good balance between convergence speed and stability.
  
  Early tests showed that the laser light source within ACE exhibited long-term centroid drift  on a timescale comparable to our response matrix collection and closed-loop tests (see next section).   Thus, the laser centroid position could be different between the response matrix calculation (i.e. the influence function) and its implementation in the spatial LDFC closed loop.   To monitor and correct (within 1 pixel) the estimate of the centroid position, we introduced a single speckle into the dark hole prior to compare the centroid position at the start of the RM calculation and that during closed-loop tests, shifting the bright and dark field pixel masks by the offset between these two centroid measurements.   Typical offsets were on the order of 2--4 pixels (0.06--0.12 $\lambda$/D); typical drift during closed-loop tests described below was on order of $\sim$ 1--2 pixels.
  %each iteration of the response matrix calculation and before 
  %implementatio

\begin{figure*}[ht]
   \begin{center}
  % \begin{tabular}{c} %% tabular useful for creating an array of images 
  %\centering
   %\includegraphics[scale=0.4,clip]{ldfc_demo_twosin_sept2019.png}
   \includegraphics[width=1.0\textwidth]{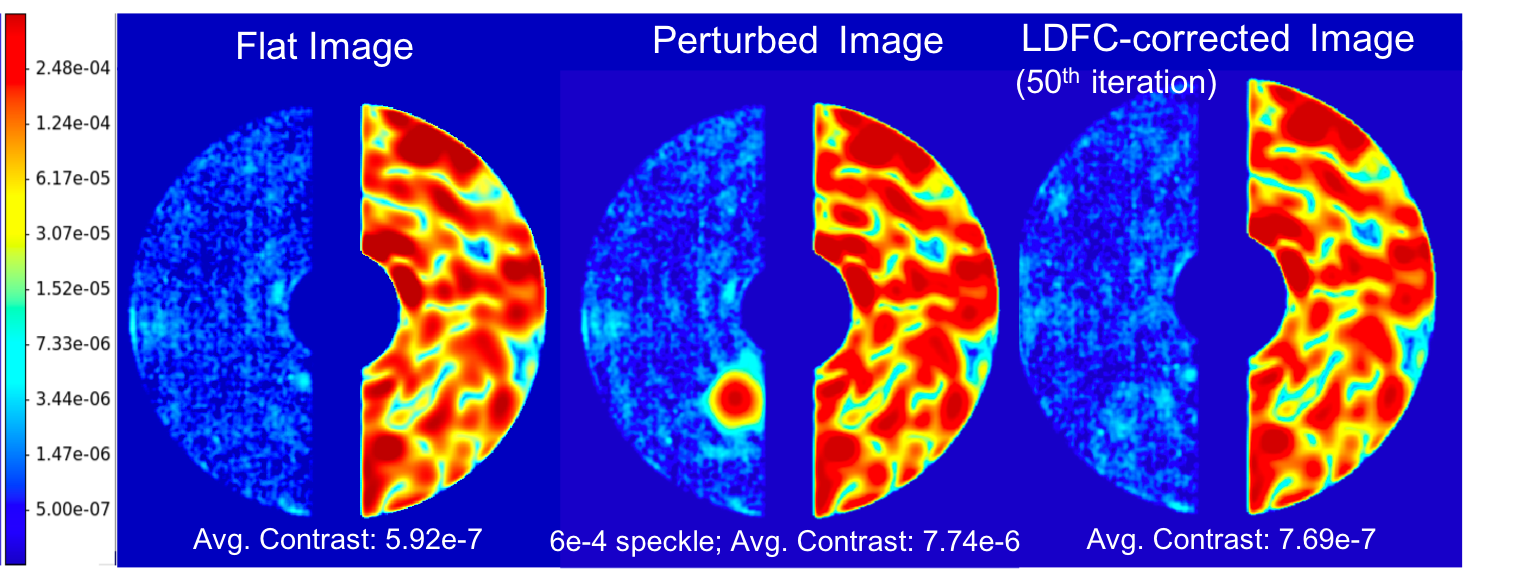}
   %\includegraphics[scale=0.47,clip]{modalresponses.png}
  % \end{tabular}
   \end{center}
   \vspace{-0.25in}
   \caption
%>>>> use \label inside caption to get Fig. number with \ref{}
   { \label{fig:ldfcdemosep2019} 
  Sequence of focal plane camera images from our 19 September 2019/``Single Speckle" experiment showing that LDFC removes a bright speckle and drives the dark field back to an average contrast within 30\% of its original value.  The spatial scale for the dark (left) and bright (right) field regions is given in Table 1 and is roughly 1.5--5.1 $\lambda$/D: regions outside this range are masked out.}
   \end{figure*}
  \subsection{Linear Dark Field Control Experimental Setup}
To test the efficacy of LDFC, for each experiment we introduce a perturbation in the pupil plane by slightly changing the DM shape from its map after the DH is created.   This phase perturbation degrades the DH, and we use the LDFC control loop to restore it.

We introduced several different types of phase perturbations that result in a range of different focal plane aberrations (Table \ref{tab:experiment}, rightmost column).   Below, we describe these perturbations and list the date on which we performed these experiments.

\begin{itemize}
    \item \textbf{1. A Single Speckle (19 September 2019)} - We introduced a sine wave perturbation on the DM to yield a bright speckle with a peak contrast of $\sim$ 2$\times$10$^{-4}$ into the DH.
    
  \item \textbf{2. A Pair of Speckles (23 December 2019)} - We introduced  sine wave perturbations on the DM to yield two bright speckles with peak contrasts of $\sim$ 5$\times$10$^{-5}$ into the DH.
        
    \item \textbf{3. Low Spatial Frequency Aberration (12 January 2020)} - To introduce this aberration, we poked a single actuator by an amplitude comparable to that used for our response matrix calculations, yielding a large region of the DH degraded to 10$^{-5}$ contrast.   

    \item \textbf{4. Complex Aberrations (25 January 2020)} - To test for LDFC's ability to correct for more complex aberrations, we introduced a weighted, linear combination of sine wave perturbations on the DM, yielding three bright speckles in the DH, each with peak contrasts greater than 10$^{-4}$, along with with fainter aberrations at the $\sim$ 10$^{-5}$ level.
\end{itemize}

For LDFC to be valuable, it must correct for speckles substantially brighter than the original DH average intensity, hold this correction for a large number of iterations, and exhibit an advantage (either in restored contrast or duty cycle) over standard DM probing methods like EFC or speckle nulling.   For Experiment \#1, we monitor the average intensity of the DH for over 100 iterations to assess whether the loop is stable.   For Experiments \#2--4, we compared LDFC's performance to that from speckle nulling.  We used the same initial DM shape, the same (to within $\sim$ 5\%) starting DH contrast, and the same perturbation.   We compared the DH contrast over 100 iterations from speckle nulling, the number of iterations needed to restore the DH, and the number of DM shape changes needed.
%get to the contrast floor, the n loop's ability to restore and sustain the dark hole over 100 iterations, 

To evaluate the efficacy of LDFC, for Experiments \#2--4 we measured the contrast over scoring regions covering the locations of the perturbations in the dark field.   For Experiments \#2--3, we selected 10$\lambda$/D squared  regions enclosing the two speckles and the peak intensity of the low spatial frequency perturbation, respectively.   For Experiment \#4, we selected the upper half of the DH (roughly 45 $\lambda$/D squared).   For these three experiments, the aberrations were adjusted to yield a factor of $\sim$ 10 or more degradation in the average contrast over the scoring region.\footnote{Key changes occurred after the September experiment.   After September, multiple bad/weak actuators appeared on the ACE deformable mirror.  In subsequent experiments, we masked these actuators in the initial creation of the DH and in the LDFC restortion of the DH.   In the September experiment, the inner radius for the DH was smaller than the inner radius of the bright field.   We found that this mismatch led to LDFC being unable to correct for perturbations closer to the optical axis.   In subsequent experiments, we matched the inner radius for both regions.  Finally, we coadded 10 times more images together in the December-January experiments to better illuminate the residual light within the DH.} 

%To test the efficacy of spatial LDFC, we stored the DM shape producing our dark hole and then introduced a grid of DM perturbations to degrade it.   Each perturbation corresponds to a single actuator poke -- as in our $RM$ calculations -- instead of a sine wave modulation producing a pair of speckles in the BF and DF or a Kolmogorov phase screen.   We set the perturbation amplitude such that it produced a factor of 2--5 degradation in the average DF intensity, intermediate between the original DF intensity and the BF intensity.   

%\textbf{Add comparison between DF and BF for delta(I)}
\begin{figure*} [ht]
  % \begin{center}
  % \begin{tabular}{c} %% tabular useful for creating an array of images 
 % \centering
   \includegraphics[width=0.997\textwidth,trim=0mm 0mm 0mm 0mm,clip]{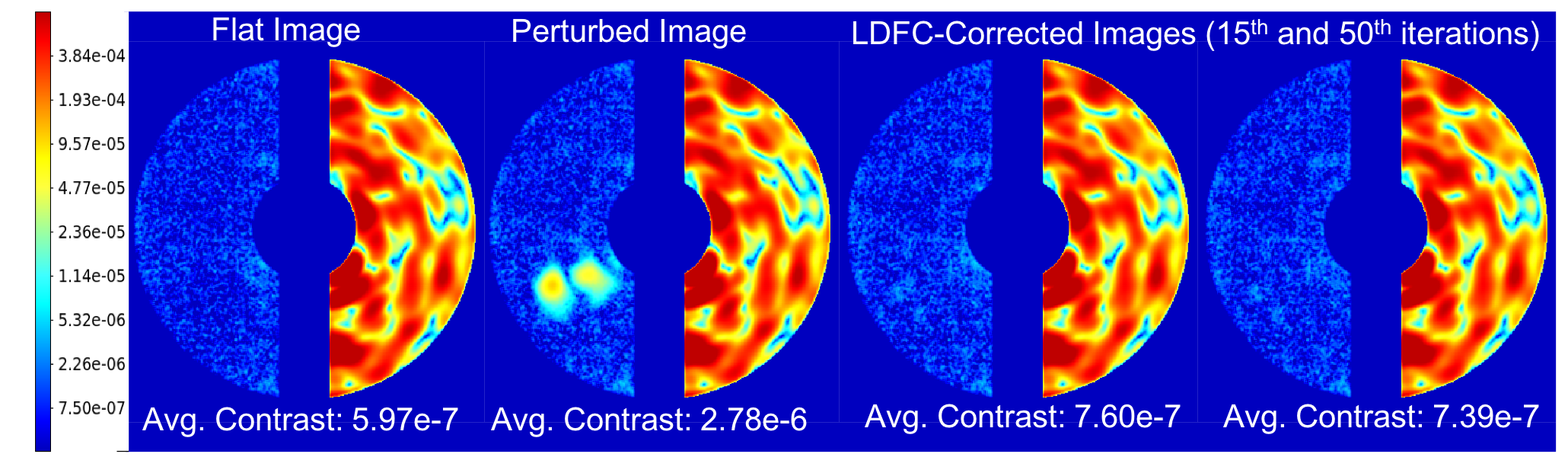}\\
    \includegraphics[width=1\textwidth]{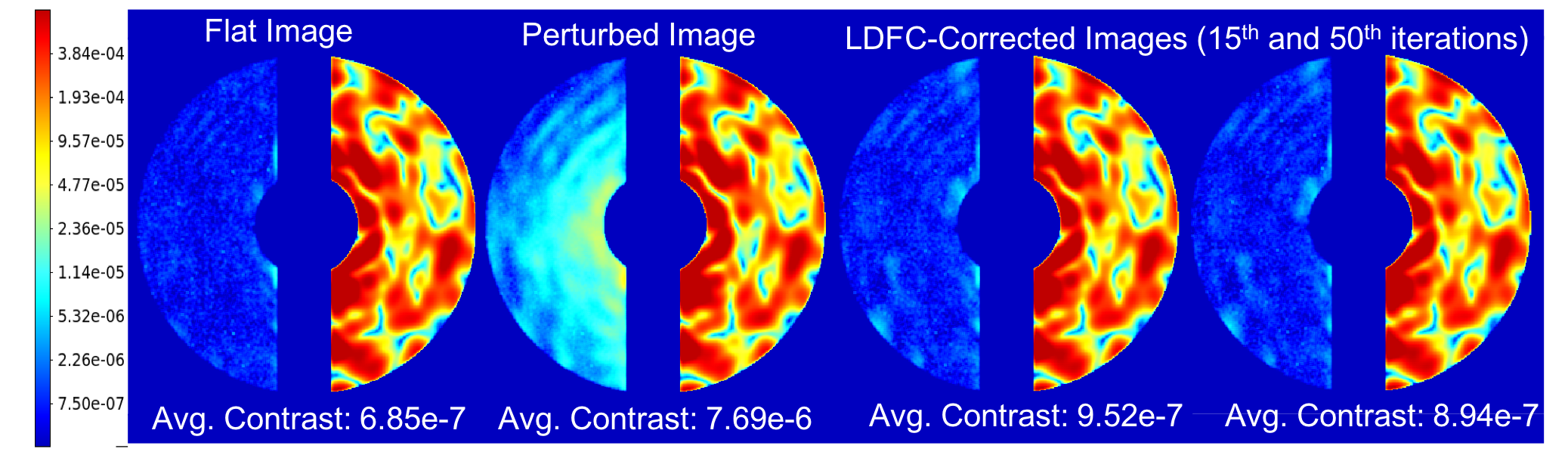}\\
      \includegraphics[width=0.999\textwidth]{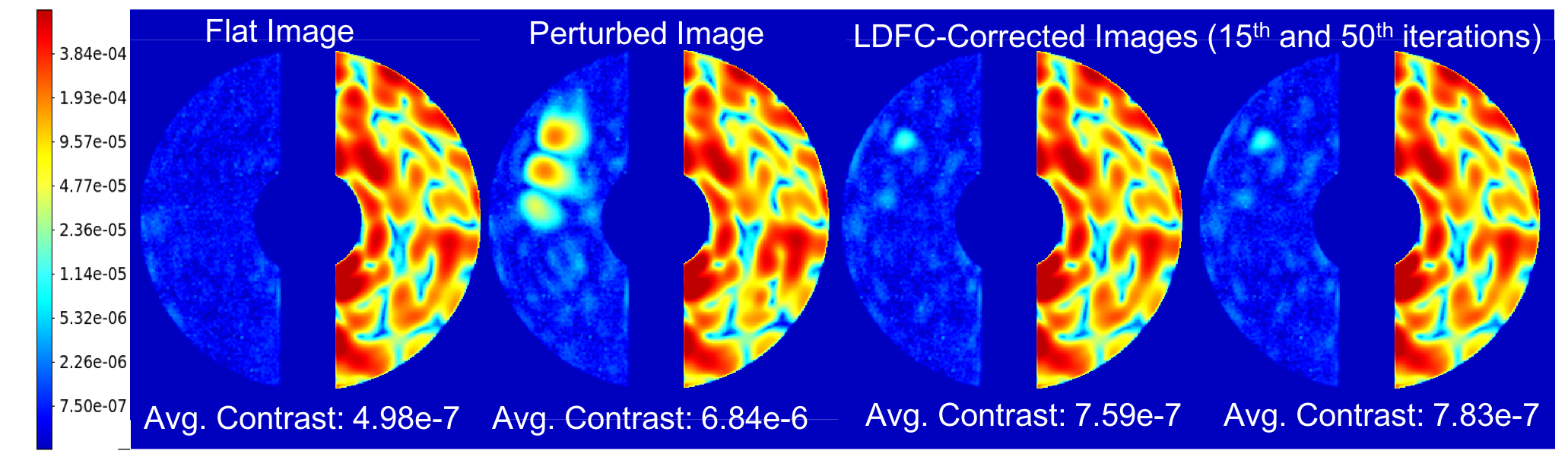}
   %\includegraphics[scale=0.47,clip]{modalresponses.png}
  % \end{tabular}
  % \end{center}
   \vspace{-0.2in}
   \caption
%>>>> use \label inside caption to get Fig. number with \ref{}
   { 
   Sequence of images for Linear Dark Field Control experiments conducted on 23 December 2019``Pair of Speckles" (top row), 12 January 2020/``Low Spatial Frequency Aberration" (middle row), and 25 January 2020/``Complex Aberration" experiments (bottom row).  Shown are the initial camera image after the creation of a dark hole (left), the camera image after the introduction of a perturbation that degrades the dark hole (middle-left), and images after the 15th and 50th iteration of Linear Dark Field Control (middle-right, right).  The spatial scale is the same as in Figure \ref{fig:ldfcdemosep2019}.  Even for strong perturbations degrading contrast by over an order of magnitude, LDFC still returns the average intensity of the dark hole to within 20-40\% of its original value.    
  % Contrast per iteration for the 23 December 2019 (left), 12 January 2020 (middle), and 25 January 2020 experiments for Linear Dark Field Control compared to performance of the speckle nulling algorithm used to create the dark hole.  Horizonal black lines denote the initial average contrast within evaluation region before a perturbation is introduced to degrade the dark hole by a factor of 10-12.   Within the dark hole, LDFC (solid magenta line) converges to within 20-40\% contrast twice as fast as deformable mirror probing with speckle nulling (blue dot-dashed line) and with a factor of 10 fewer deformable mirror commands (blue long-dashed line).   
 }
 \label{fig:ldfcseqimages_2} 
   \end{figure*}
\section{Results}

%Figures \ref{fig:ldfcdemosep2019}--\ref{fig:ldfccontrastperiteration3} display the results of our four experiments. 
%\subsection{Demonstration of Spatial Linear Dark Field Control}
\subsection{Spatial LDFC Dark Hole Restoration}
%Figure \ref{fig:ldfcdemosep2019} demonstrates the performance of Spatial Linear Dark Field Control from September 2019 experiment.    

Figures 
%\ref{fig:ldfcdemosep2019}, \ref{fig:ldfcseqimages_2}, \ref{fig:ldfcrangesept19}, \ref{fig:ldfccontrastperiteration}, \ref{fig:ldfccontrastperiteration2}, and
%\ref{fig:ldfccontrastperiteration3} 
\ref{fig:ldfcdemosep2019} and \ref{fig:ldfcseqimages_2}
%presents a sequences of images from Experiments 2--4, illustrating 
illustrate the ability of Spatial LDFC to restore a DH corrupted by a range of different phase-induced aberrations: one bright speckle, two speckles, a broad low spatial frequency aberration, and three speckles with fainter second order peaks.   For the September 2019, experiment, an initial (`flat') image with an average DH contrast of $\sim$ 6$\times$10$^{-7}$ is degraded by a single speckle with a peak contrast a factor of 1000 larger.   The average contrast over the entire DH increases by a factor of $\sim$ 11.    Spatial LDFC immediately begins removing this speckle.  LDFC achieves a restored DH with a contrast within $\sim$ 30\% of the original DH average intensity (righthand panel of Figure \ref{fig:ldfcdemosep2019}, Figure \ref{fig:ldfcrangesept19}).  
For the December and January experiments, the initial DH contrasts range between 5 and 6.9$\times$10$^{-7}$ over the entire DH and 4.6--5.9$\times$10$^{-7}$ within the relevant scoring regions (Figure \ref{fig:ldfccontrastperiteration}).   Aberrations degrade the DH by a factor of 4.6--13.7; within the scoring regions, the DH contrast is made 13-26 times brighter to $C$ $\sim$ 7.2$\times$10$^{-6}$--1.2$\times$10$^{-5}$.  

Spatial LDFC then reduces the aberrated DH contrast by a factor of 3.7--9 over the entire field (7.4--9.5$\times$10$^{-7}$; left two panels of Figure \ref{fig:ldfcseqimages_2}) and a factor of 10--13 over the scoring regions (7.4--9.3$\times$10$^{-7}$; Figure \ref{fig:ldfccontrastperiteration}).   Over the entire DH region, LDFC reaches an average contrast within a factor of 1.2--1.4 of the pre-aberrated state.   Within the scoring region, the restored average contrast is within a factor of 1.2--1.7 of its original value.

For Experiments 1--3, the initial aberration is (almost) perfectly removed by LDFC and most residual left by LDFC is largely confined to the edges of the DH region.   We speculate that LDFC does not fully remove residual signal because a) regions near the edge of the dark field/bright field are generally more difficult to correct and b) the correction becomes 'noisier' as average contrast approaches the read noise level.     For Experiment $\#$4, the initial aberration is largely removed but a faint residual core ($\sim$ 7 pixels in radius) of the brightest speckle remains after LDFC at a 10$^{-5}$ level\footnote{Section \ref{sec:discussion} discusses potential reasons why this residual signal and that in other experiments remain}.          

\subsection{Stability of Spatial LDFC Dark Hole Correction and Comparison to Speckle Nulling}

\begin{figure} [ht]
  % \begin{center}
  % \begin{tabular}{c} %% tabular useful for creating an array of images 
  %\centering
   \includegraphics[width=0.5\textwidth,trim = 12mm 70mm 0mm 75mm, clip]{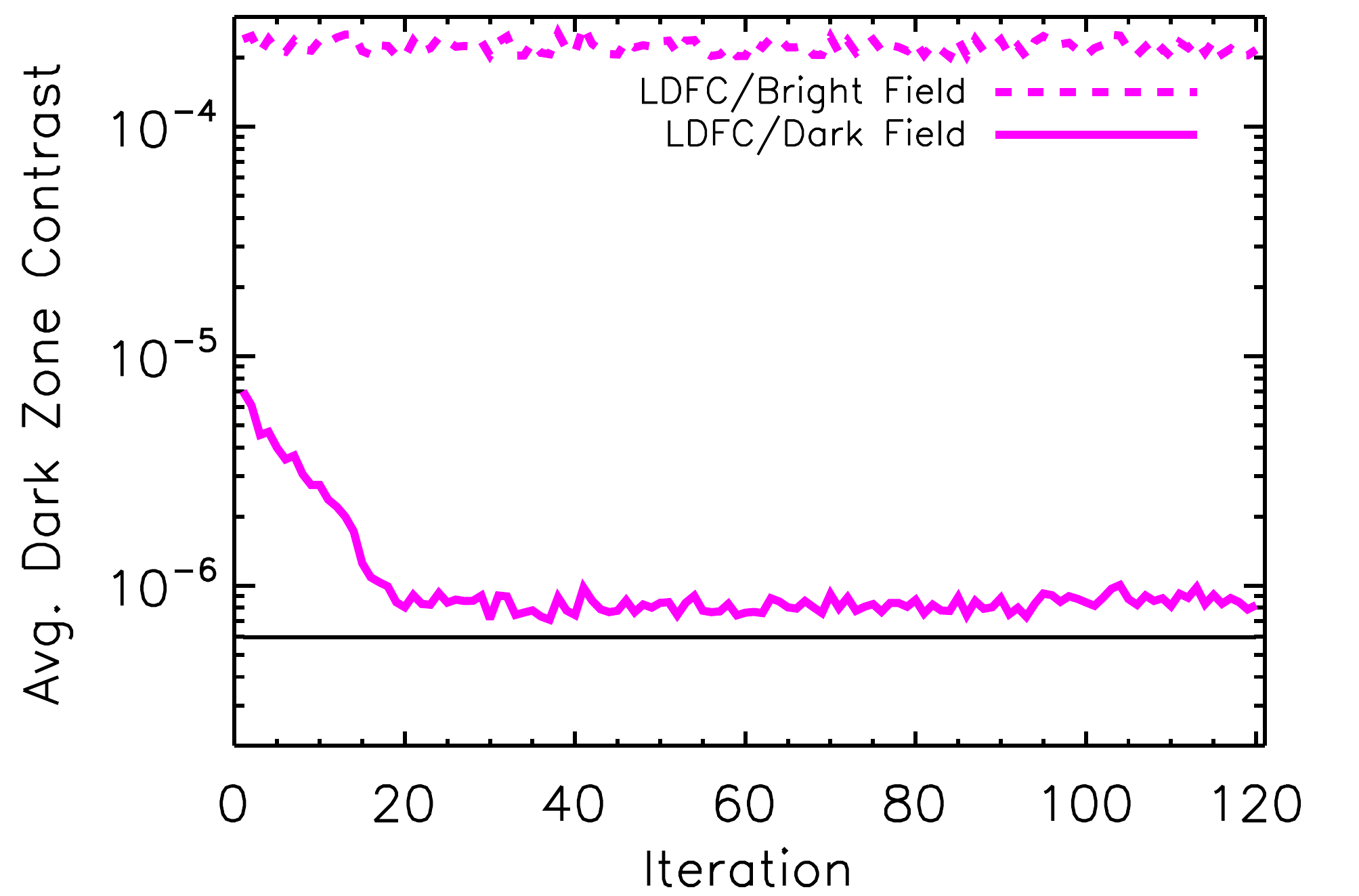}
   %   \includegraphics[width=0.5\textwidth,trim = 12mm 5mm 0mm 5mm, clip]{avgsignal_sine1_longtest.pdf}
   % \includegraphics[width=0.45\textwidth,trim = 10mm 5mm 10mm 5mm,clip]{Figures/avgsignal_single_actuator_test_919.eps}
   %   \includegraphics[width=0.32\textwidth,trim = 10mm 5mm 10mm 5mm,clip]{Figures/ldfc0125loop.eps}
   %\includegraphics[scale=0.47,clip]{modalresponses.png}
  % \end{tabular}
  % \end{center}
   \vspace{-0.2in}
   \caption
%>>>> use \label inside caption to get Fig. number with \ref{}
   { \label{fig:ldfcrangesept19} 
   Analysis of our 19 September 2019/``Single Speckle" experiment.   Contrast per iteration for LDFC for the perturbation introduced in Figure \ref{fig:ldfcdemosep2019}, showing that LDFC sustains a dark hole below 10$^{-6}$ contrast for over 100 consecutive iterations. 
   %(right) Relative contrast between the unperturbed dark field and LDFC after first introducing four large-amplitude separate single-actuator deformable mirror perturbations.  After 15 iterations, LDFC is still converging to near the original dark hole contrast. 
   %LDFC exhibits a slower, shallower convergence when actuator number 1053 is perturbed, likely because the perturbation moves the system response (nearly) out of the linear response regime.
 }
   \end{figure}

\begin{figure*}
   \begin{center}
  % \begin{tabular}{c} %% tabular useful for creating an array of images 
  %\centering
   \includegraphics[width=0.369\textwidth,trim = 12.8mm 70mm 11.7mm 70mm, clip]{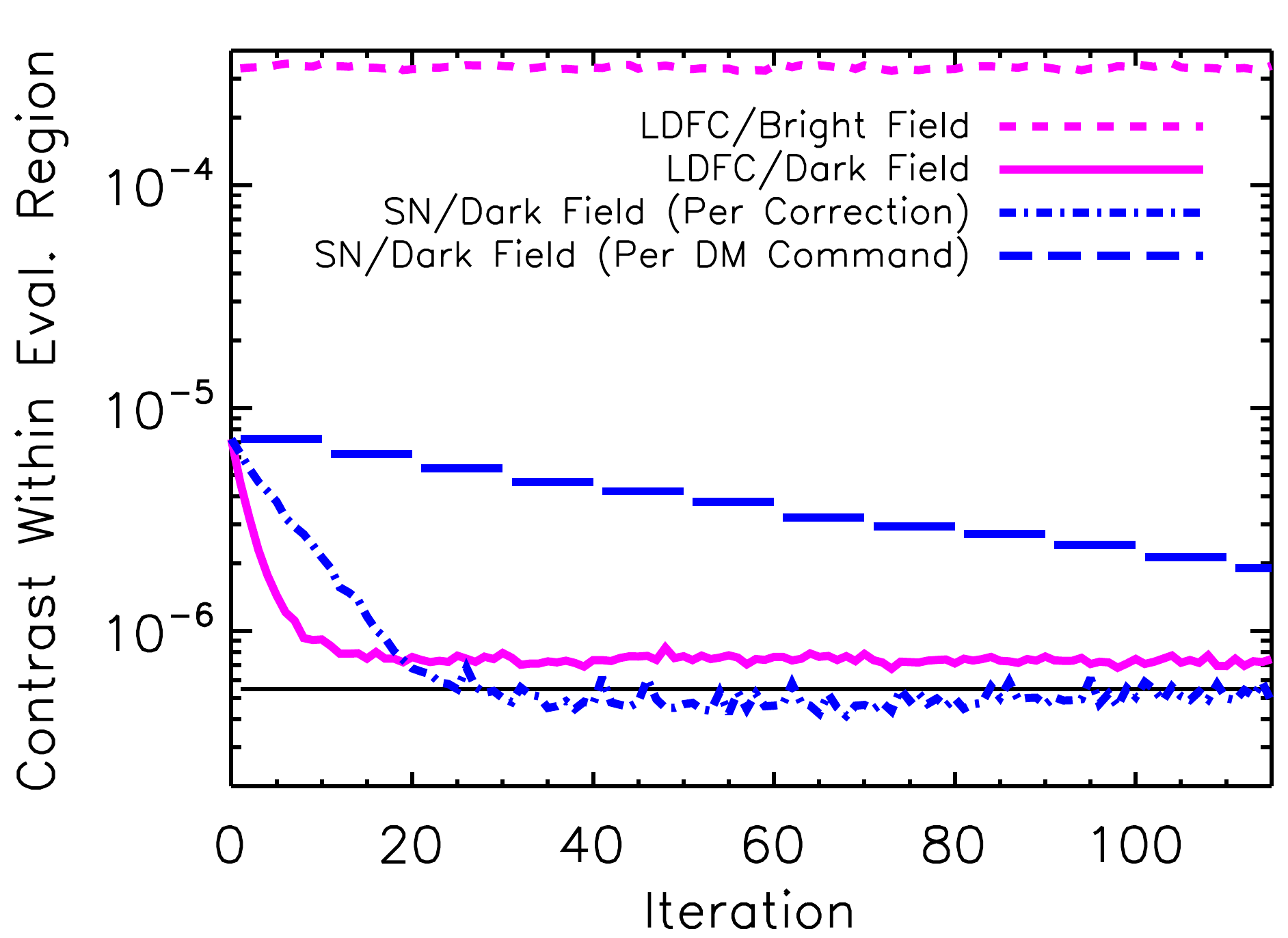}
      \includegraphics[width=0.30\textwidth,trim = 37.2mm 5mm 11.7mm 0mm,clip]{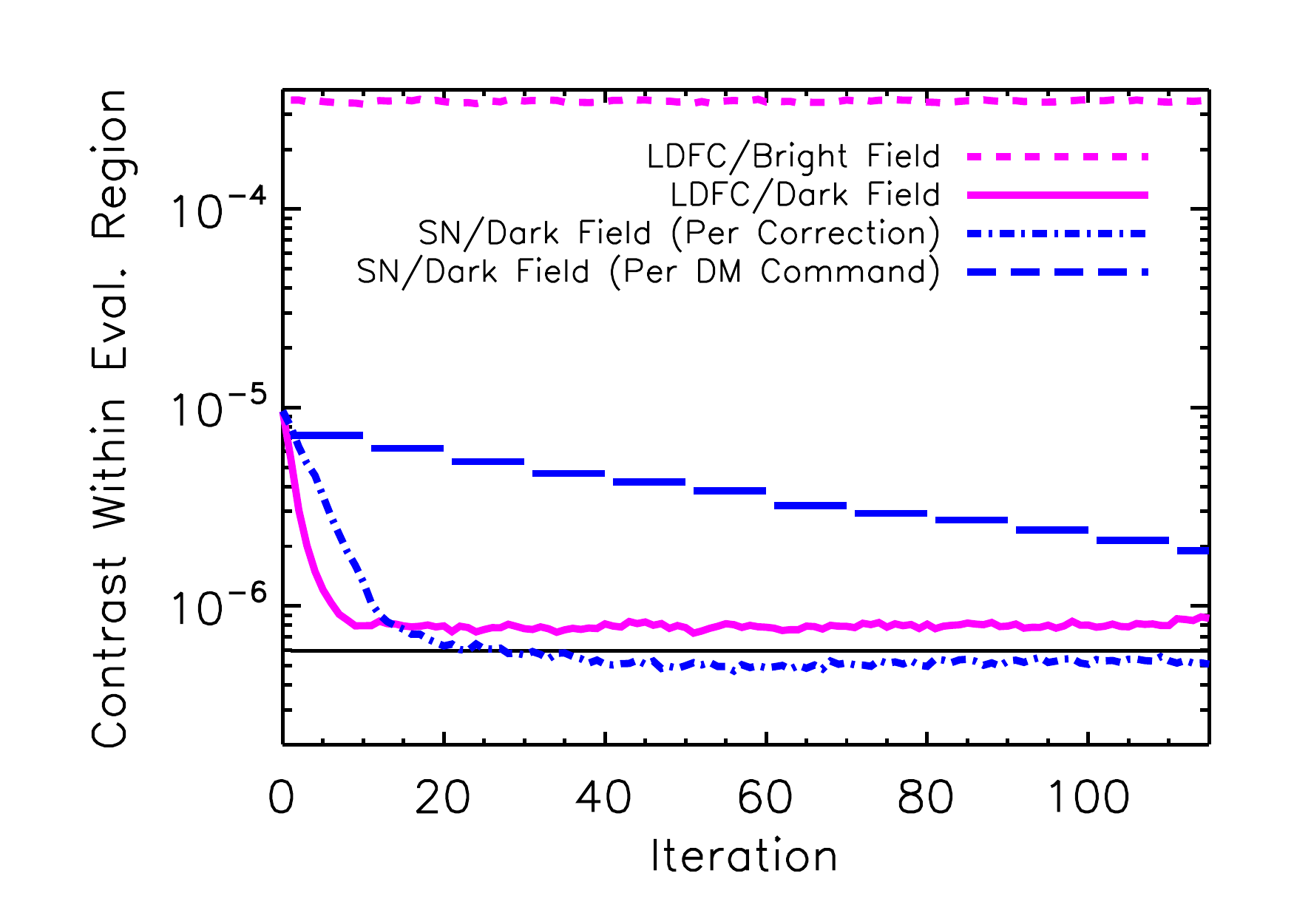}
         \includegraphics[width=0.305\textwidth,trim = 45mm 68mm 11.7mm 70mm,clip]{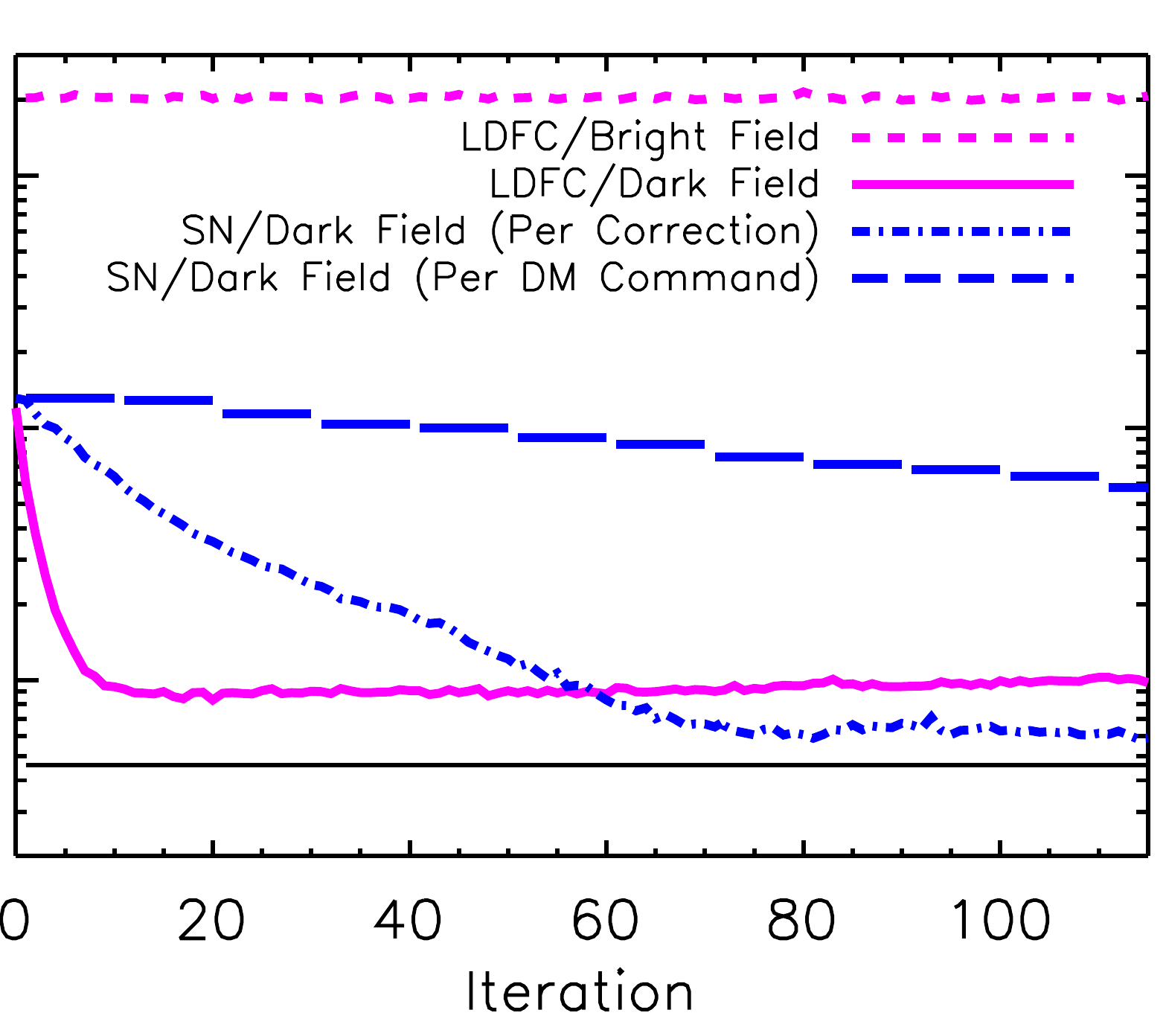}
      %      \includegraphics[width=0.369\textwidth,trim = 12.8mm 5mm 11.7mm 10mm, clip]{ldfc1223loop.pdf}
      %\includegraphics[width=0.3095\textwidth,trim = 37.2mm 5mm 11.7mm 10mm,clip]{ldfc0112loop.pdf}
        % \includegraphics[width=0.3095\textwidth,trim = 37.2mm 5mm 11.7mm 10mm,clip]{ldfc0125loop.pdf}
   % \includegraphics[width=0.32\textwidth,trim = 10mm 5mm 10mm 5mm,clip]{Figures/ldfc0112loop.eps}
   %   \includegraphics[width=0.32\textwidth,trim = 10mm 5mm 10mm 5mm,clip]{Figures/ldfc0125loop.eps}
   %\includegraphics[scale=0.47,clip]{modalresponses.png}
  % \end{tabular}
   \end{center}
   \vspace{-0.2in}
   \caption
%>>>> use \label inside caption to get Fig. number with \ref{}
   { \label{fig:ldfccontrastperiteration} 
   Contrast per iteration for the 23 December 2019/``Pair of Speckles" (left), 12 January 2020/``Low Spatial Frequency Aberration" (middle), and 25 January 2020/``Complex Aberration" (right) experiments for Linear Dark Field Control compared to performance of the speckle nulling algorithm used to create the dark hole.  Horizonal black lines denote the initial average contrast within evaluation region before a perturbation is introduced to degrade the dark hole by a factor of 10-12.   Within the dark hole, LDFC (solid magenta line) converges to within 20-40\% contrast twice as fast as deformable mirror probing with speckle nulling (blue dot-dashed line) and with a factor of 20 or  fewer deformable mirror commands (blue long-dashed line).   
 }
   \end{figure*}

The Spatial LDFC-restored DH shows long-term stability.  For the September experiment, the DH contrast converges after 18 iterations to a value of 8$\times$10$^{-7}$ $\pm$ 6.5$\times$10$^{-8}$ for the next 105 iterations (Figure \ref{fig:ldfcrangesept19}).   The bright field stays constant within about the same fractional value: the average intensity fluctuations are expected given measured varations of the laser brightness with time ($\sim$ 5\%).   For the December and January experiments (Figure \ref{fig:ldfccontrastperiteration}), convergence to a final (largely-)restored DH occurs within 5-10 iterations and stays constant within 10\% for 110 iterations.

LDFC shows evidence for significantly improved efficiency compared to DM probing methods like speckle nulling.  Speckle nulling is able to restore the DH to a contrast level $\sim$ 5--6$\times$10$^{-7}$: 25--40\% lower than LDFC and comparable to the inital, unperturbed DH contrast.   However, speckle nulling requires 20--70 iterations to reach its final contrast level (dash-dotted blue lines).   Reaching the contrast level achieved by LDFC requires a factor of 2--5 more iterations.  

When analyzed in terms of DM commands, the efficiency advantage of LDFC is significantly larger.   For each iteration, speckle nulling requires multiple DM probes in order to estimate the phase of residual speckles in the dark zone and estimate amplitude: 10 for our implementation.   Speckle nulling requires a factor of 20--50 more DM commands to reach the contrasts achieved by LDFC.   The advantage in duty cycle is particularly large for complex aberrations introduced into the focal plane (Experiment 4).
%\subsection{Comparison to Speckle Nulling}
%Compare to SN ....  SN coverages back exactly to DH because experiment set up DH to be bright enough to directly sense.   But still took factor of 2-5 longer number of iterations.   While LDFC one DM command per iteration, SN requires 5 DM probes to get correction.   Per DM command, LDFC factor of 10-25 more efficient.   Advantage most pronounced for more complex speckle patterns not replicatable with simple DM shapes.

\section{Discussion\label{sec:discussion}}
\subsection{Summary of Results and Implications}
This study presents the first laboratory demonstrations of Spatial Linear Dark Field Control at contrast levels ($\sim$5$\times$10$^{-7}$) and separations ($\sim$ 1.2--5.2 $\lambda$/D) approaching the raw performance needed to image some jovian planets in reflected light around the nearest Sun-like stars with space-borne coronagraphic instruments like WFIRST-CGI and with ELTs around low-mass stars.   In four experiments conducted with the ACE testbed, we introduced a range of different phase perturbations that degraded the average intensity of the dark hole (a $\ge$ (10$\lambda/D$)$^{2}$ scoring area within the dark hole) by a factor of 5--10 (13-26).   Spatial LDFC restores the average intensity of the dark hole to within a factor of 1.2--1.4 of its original contrast.   In the scoring region focused on the perturbations, Spatial LDFC converges to within a factor of 1.2--1.7 of the original dark hole contrast.   Spatial LDFC maintains the average dark hole contrast for over 100 iterations.

The Spatial LDFC experiments demonstrate significant potential advantages for maintaining a dark hole over methods that use DM probing to directly remove aberrations in the DH.   When presented with the same aberrations, speckle nulling is able to achieve 25--40\% deeper contrasts than LDFC.   However, speckle nulling requires a factor of 2--5 more iterations to match Spatial LDFC's performance.   As speckle nulling requires multiple modulations per iteration to estimate the phase of residual speckles in the DH, the duty cycle advantage for LDFC in terms of DM commands is substantial: a factor of 20--50 in our experiments. 

Linear Dark Field Control may provide a promising path forward to maintain dark holes without relying on DM modulation and probing, especially if its small performance gap compared to probing techniques is closed and if possible null space can be mitigated.    The far shorter duty cycle offered by LDFC improves the efficiency of high-contrast imaging observations. 
  By construction, LDFC drives the dark hole back to its initial state, which should improve the temporal correlation of speckles, while the LDFC loop is in operation.   A shorter duty cycle and increased dark hole stability should substantially improve our ability to image mature solar system-like planets in reflected light over the next two decades.
\subsection{Experiment Drawbacks and Null Space with Spatial LDFC}
In our experiments, Spatial LDFC's main drawback is that it converges to a dark hole contrast a factor of 1.2--1.7 higher than in the pre-aberrated state.   Experimental conditions may account for much of this performance gap.   For example, laser centroid drift during the RM calculation may compromise the accuracy of our influence function for LDFC.   Drift during the closed-loop tests themselves likewise limits the accuracy of our correction.   Instability in the laser power on the few percent level may limit accuracy as the average DH contrast approaches the initial, pre-aberrated state.  The \textit{G}-matrix encoding relationship between DM pokes and complex amplitudes for EFC may change with time, and similarly the RM for LDFC degrades with time.
Weak/bad actuators on the DM not currently flagged may lead to a poor influence function determination and impede convergence. 

Some of the aberrations degrading the DH may produce intensity variations at/near lowest-flux regions of the bright field may lie also in a quadratic response regime.  A region of the bright field in the quadratic response regime would preclude identifying a unique DM shape that could be applied to restore its initial state and that of the dark field.   This is a particularly relevant possibility for the residual speckle core left in Experiment \#4, as bright field region 180 deg from that speckle is at a local minimum in flux.

Laser drift can be better corrected by monitoring the centroid position during the RM calculation and by improving our loop speed.  Better regularization can limit the impact of laser instability.   Future Spatial LDFC experiments at ACE will be conducted with a repaired DM or a replacement free of bad/dead actuators and with a more efficient loop to reduce the impact of system RM evolution.  

A key concern for future progress with Spatial LDFC is the existence of null space, where a given pupil-plane perturbation aberrates the dark field but produces a negligible change in the bright field.   Null space is expected to include a combination of amplitude and phase errors, which can create single-side speckles.   By construction, our experiments only demonstrated spatial LDFC's ability to remove phase errors that be represented by a linear combination of DM pokes (e.g. not perturbations with a spatial frequency higher than the DM pitch).  More importantly, we did not, in this experiment, introduce amplitude errors in the pupil plane.  Amplitude errors can result from reflectivity variations in system optics and phase-induced errors due to out of plane optics.  Amplitude errors are expected to be equally important at raw contrasts in the range of 10$^{-7}$--10$^{-9}$ or below \citep[e.g.][J. Krist. pvt. comm.]{ShaklanGreen2006,Pueyo2007,Bailey2018}.  

Null space can be addressed in the following ways.  To partially compensate for null space for Spatial LDFC,  the bright field mask could be adjusted, adding pixels exterior to but on the same side as the dark field, to be sensitive to at least some amplitude errors.   It may be possible to treat amplitude and high spatial frequency phase perturbations by solving for an aberration map informed by a regression procedure.   Constructing such a map requires quantitative modeling of the DM and coronagraph optical train and will be the subject of future work in simulations and on the ACE testbed.  Finally, spectral LDFC \citep{Guyon2017} utilizing out-of-band measurements over the same focal plane region for the bright and dark fields, instead of different regions as in spatial LDFC, should be sensitive to both phase and amplitude aberrations provided that the main wavefront change is due to optical path difference in/near the pupil plane (where
largest optics are). Improvements in the experimental setup for LDFC will enable a better tests of the method's fundamental limits.  Masking of lower-flux pixels with the bright field can focus LDFC on focal plane regions responding linearly to perturbations.
%Alternatively, a substantial region of the bright field could be in the quadratic, not linear, response regime: different perturbations that corrupt the dark field could aberrate the bright field in the same way. 
%Generally speaking, we do not yet identify any null space, at least at the $\sim$ 10$^{-7}$ contrast level.   The potential exception to this rule is 
%The residual speckle signal left in Experiment \#4 could be an example of this behavior, as the bright field region directly opposite of this speckle covers a local minimum with a contrast of $\sim$ 10$^{-6}$ but particularly large signals at adjacent bright field pixels. Laser centroid drift and RM evolution could also potentially impede Spatial LDFC's ability to remove this signal.   

%true solar system analogues in the next two decades. 

\subsection{Future Tests of Linear Dark Field Control}
Future experiments at the ACE testbed will further mature Spatial Linear Dark Field Control.  Realistic aberrations (e.g. linear combination of Zernike modes) introduced mimicking those expected in flight for missions like WFIRST-CGI may provide a better practical test of Spatial LDFC.   Our tests focus on sudden introductions of large-intensity perturbations into the dark field.   An alternate test where smaller perturbations are periodically introduced and then corrected may better simulate closed-loop operations.  Our experiments were conducted with the residual DH signal is well illuminated.  ``Blind" tests -- where the DH residual intensity is comparable to the detector noise level over the WFS sampling time -- can better assess LDFC's advantage over DM probing techniques in the (dark field) photon-starved regime.  Adopting more advanced focal-plane wavefront sensing techniques such as EFC, Kalman filtering, or variants that optimize DM probing and integration time \citep{Groff2013,Groff2016,Sun2020} instead of speckle nulling may provide a more robust assessment of LDFC's advantages to state-of-the-art DM probing FPWFC methods.

%\textbf{REMAINING STUFF}
Upcoming/proposed NASA missions capable of imaging exoplanets in reflected light like WFIRST-CGI, HabEx, and LUVOIR require sustained raw contrasts of 10$^{-9}$--10$^{-10}$.   Vacuum chamber experiments on the \textit{High-Contrast Imaging Testbed} at the Jet Propulsion Laboratory will provide a first test of Linear Dark Field Control's efficacy at these extreme contrast regimes.   For these tests, we will employ Spectral LDFC \citep{Guyon2017}, where out-of-band focal-plane images at wavelengths bracketing that of the main sciene bandpass will be needed to restore and freeze the dark hole.  Typical exposures for these missions will several to tens of hours.   A key milestone then will be to demonstrate stability at $<$ 10$^{-9}$--10$^{-10}$ contrast for tens of hours.

\subsection{Practical Implementation of Linear Dark Field Control with WFIRST-CGI and Extremely Large Telescopes}

For implementation of Spatial LDFC either on the ground or (especially) in space, a key challenge will be the extremely high dynamic range required at the focal plane.  The residual signal within the dark field must be illuminated and the bright field (10$^{3}$--10$^{4}$ times brighter) must be in the linear response regime.  A neutral density filter covering the bright field or differential readout of the bright and dark regions could work around this problem. 

Utilizing a version of Linear Dark Field Control may be possible with upcoming NASA missions, in particular WFIRST-CGI, but with some adjustments.   WFIRST-CGI uses two deformable mirrors to generate a 360$^{o}$ dark hole.  Nominally, Spatial LDFC would be at-best modestly effective for this setup as regions opposite a one-sided dark hole sample the same spatial scale, while regions exterior to the dark field sample a different scale and are not necessarily fully sensitive to the same focal-plane aberrations.
%\footnote{Preliminary tests at ACE appear to support this conclusion.}.

To compensate for such a setup and still utilize Spatial LDFC, one possibility is to simply create a one-sided dark hole and use the opposite side for wavefront sensing.   As WFIRST-CGI's technical demonstrations and foreseable, subsequent science observations would focus on previously-identified exoplanets with known positions \cite[e.g.][]{GrecoBurrows2015}, the dark hole region could be chosen beforehand with only a modest increase in the contrast requirements due to finite element corrections \citep{Mawet2014}.

The descope of the WFIRST-CGI integral field spectrograph prevents the utilization of Spectral LDFC as originally conceived of in \citet{Guyon2017}; simultaneous, out-of-band broadband filter observations at wavelengths bracketing that of a science observation likewise are not possible.   However, an ``open loop" version of Spectral LDFC could be adopted, consisting of periodic images in out-of-band filters to ``touch up" the wavefront correction.  While less efficient than standard Spectral LDFC, the duty cycle for this method would be far smaller than the nominal strategy of slewing back to a PSF reference star to reestablish the dark hole.   Spectral LDFC samples the same region of the image plane, in and out of band.  As the chief challenge with any flavor of LDFC going forward is null space between the dark field and bright field dominated by amplitude and phase-induced amplitude errors, spectral LDFC could potentially circumvent null space limitations of Spatial LDFC.

Utilization of LDFC with ELTs should be more straightforward.   For instance, the \textit{Planetary Systems Imager} on the \textit{Thirty Meter Telescope} envisions high-contrast imaging observations with an integral field spectrograph that covers 0.6--1.8 $\mu m$ \citep{Fitzgerald2019}.   For either a one-sided dark hole or a coronagraph yielding a deep correction over a small (e.g. 10\%) bandpass, a version of LDFC could be employed.   At raw contrasts needed to image Earth-like planets around nearby low-mass stars (10$^{-6}$ at 1 $\mu m$), phase errors still dominate the wavefront error budget \citep{Guyon2018a}.

A key challenge for the ground will be to demonstrate that the LDFC loop can converge significantly faster than the coherence time of atmospheric speckles: $t_{\rm o}$ $\sim$ 5--10 $ms$, even for the most high-contrast imaging-friendly site, Maunakea.     To mature LDFC for these purposes, we plan to implement it as a separate wavefront control loop with the Subaru Coronagraphic Extreme Adaptive Optics project \citep{Jovanovic2015,Lozi2018,Currie2019a}.  Internal source tests with SCExAO using a turbulence simulator already show promise (Miller \& Bos, in prep.).

\textbf{Acknowledgements} -- 
We thank the NASA Strategic Astrophysics Technology (SAT) Program for their generous support of this project (Grant $\#$ 80NSSC19K0121); T.C. is supported by a NASA Senior Postdoctoral Fellowship.    ExoTAC members provided extremely helpful guidance for defining our Milestone $\#$1 requirements, which are reported in this publication.  Mamadou N'Diaye, John Krist, and the anonymous referee provided helpful comments.
 
%Finally, T.C. thanks Mengshu Xu for her exceptional patience and understanding 
 \bibliography{ldfc_pasp} % bibliography data in report.bib

\begin{thebibliography}{}
\expandafter\ifx\csname natexlab\endcsname\relax\def\natexlab#1{#1}\fi
\providecommand{\url}[1]{\href{#1}{#1}}
\providecommand{\dodoi}[1]{doi:~\href{http://doi.org/#1}{\nolinkurl{#1}}}
\providecommand{\doeprint}[1]{\href{http://ascl.net/#1}{\nolinkurl{http://ascl.net/#1}}}
\providecommand{\doarXiv}[1]{\href{https://arxiv.org/abs/#1}{\nolinkurl{https://arxiv.org/abs/#1}}}

\bibitem[{{Bailey} {et~al.}(2018){Bailey}, {Bottom}, {Cady}, {Cantalloube}, {de
  Boer}, {Groff}, {Krist}, {Millar-Blanchaer}, {Vigan}, {Chilcote}, {Choquet},
  {De Rosa}, {Girard}, {Guyon}, {Kern}, {Lagrange}, {Macintosh}, {Males},
  {Marois}, {Meshkat}, {Milli}, {N'Diaye}, {Ngo}, {Nielsen}, {Rhodes}, {Ruane},
  {van Holstein}, {Wang}, \& {Xuan}}]{Bailey2018}
{Bailey}, V.~P., {Bottom}, M., {Cady}, E., {et~al.} 2018, in Society of
  Photo-Optical Instrumentation Engineers (SPIE) Conference Series, Vol. 10698,
  \procspie, 106986P, \dodoi{10.1117/12.2313820}

\bibitem[{{Barman} {et~al.}(2015){Barman}, {Konopacky}, {Macintosh}, \&
  {Marois}}]{Barman2015}
{Barman}, T.~S., {Konopacky}, Q.~M., {Macintosh}, B., \& {Marois}, C. 2015,
  \apj, 804, 61, \dodoi{10.1088/0004-637X/804/1/61}

\bibitem[{{Beichman} {et~al.}(2010){Beichman}, {Krist}, {Trauger}, {Greene},
  {Oppenheimer}, {Sivaramakrishnan}, {Doyon}, {Boccaletti}, {Barman}, \&
  {Rieke}}]{Beichman2010}
{Beichman}, C.~A., {Krist}, J., {Trauger}, J.~T., {et~al.} 2010, Publications
  of the Astronomical Society of the Pacific, 122, 162, \dodoi{10.1086/651057}

\bibitem[{{Belikov} {et~al.}(2011){Belikov}, {Pluzhnik}, {Witteborn}, {Greene},
  {Lynch}, {Zell}, \& {Guyon}}]{Belikov2011}
{Belikov}, R., {Pluzhnik}, E., {Witteborn}, F.~C., {et~al.} 2011, Society of
  Photo-Optical Instrumentation Engineers (SPIE) Conference Series, Vol. 8151,
  {Laboratory demonstration of high-contrast imaging at inner working angles 2
  {\ensuremath{\lambda}}/D and better}, 815102, \dodoi{10.1117/12.894201}

\bibitem[{{Belikov} {et~al.}(2009){Belikov}, {Pluzhnik}, {Connelley},
  {Witteborn}, {Lynch}, {Cahoy}, {Guyon}, {Greene}, \&
  {McKelvey}}]{Belikov2009}
{Belikov}, R., {Pluzhnik}, E., {Connelley}, M.~S., {et~al.} 2009, Society of
  Photo-Optical Instrumentation Engineers (SPIE) Conference Series, Vol. 7440,
  {First results on a new PIAA coronagraph testbed at NASA Ames}, 74400J,
  \dodoi{10.1117/12.826772}

\bibitem[{{Belikov} {et~al.}(2012){Belikov}, {Pluzhnik}, {Witteborn}, {Greene},
  {Lynch}, {Zell}, {Schneider}, {Guyon}, \& {Tenerelli}}]{Belikov2012}
{Belikov}, R., {Pluzhnik}, E., {Witteborn}, F.~C., {et~al.} 2012, in Society of
  Photo-Optical Instrumentation Engineers (SPIE) Conference Series, Vol. 8442,
  \procspie, 844209, \dodoi{10.1117/12.927218}

\bibitem[{{Bord{\'e}} \& {Traub}(2006)}]{Borde2006}
{Bord{\'e}}, P.~J., \& {Traub}, W.~A. 2006, \apj, 638, 488,
  \dodoi{10.1086/498669}

\bibitem[{{Cady} {et~al.}(2016){Cady}, {Prada}, {An}, {Balasubramanian},
  {Diaz}, {Kasdin}, {Kern}, {Kuhnert}, {Nemati}, {Poberezhskiy}, {Eldorado
  Riggs}, {Zimmer}, \& {Zimmerman}}]{Cady2016}
{Cady}, E., {Prada}, C.~M., {An}, X., {et~al.} 2016, Journal of Astronomical
  Telescopes, Instruments, and Systems, 2, 011004,
  \dodoi{10.1117/1.JATIS.2.1.011004}

\bibitem[{{Chauvin} {et~al.}(2017){Chauvin}, {Desidera}, {Lagrange}, {Vigan},
  {Gratton}, {Langlois}, {Bonnefoy}, {Beuzit}, {Feldt}, {Mouillet}, {Meyer},
  {Cheetham}, {Biller}, {Boccaletti}, {D'Orazi}, {Galicher}, {Hagelberg},
  {Maire}, {Mesa}, {Olofsson}, {Samland}, {Schmidt}, {Sissa}, {Bonavita},
  {Charnay}, {Cudel}, {Daemgen}, {Delorme}, {Janin-Potiron}, {Janson},
  {Keppler}, {Le Coroller}, {Ligi}, {Marleau}, {Messina}, {Molli{\`e}re},
  {Mordasini}, {M{\"u}ller}, {Peretti}, {Perrot}, {Rodet}, {Rouan}, {Zurlo},
  {Dominik}, {Henning}, {Menard}, {Schmid}, {Turatto}, {Udry}, {Vakili}, {Abe},
  {Antichi}, {Baruffolo}, {Baudoz}, {Baudrand}, {Blanchard}, {Bazzon}, {Buey},
  {Carbillet}, {Carle}, {Charton}, {Cascone}, {Claudi}, {Costille}, {Deboulbe},
  {De Caprio}, {Dohlen}, {Fantinel}, {Feautrier}, {Fusco}, {Gigan}, {Giro},
  {Gisler}, {Gluck}, {Hubin}, {Hugot}, {Jaquet}, {Kasper}, {Madec}, {Magnard},
  {Martinez}, {Maurel}, {Le Mignant}, {M{\"o}ller-Nilsson}, {Llored}, {Moulin},
  {Orign{\'e}}, {Pavlov}, {Perret}, {Petit}, {Pragt}, {Puget}, {Rabou},
  {Ramos}, {Rigal}, {Rochat}, {Roelfsema}, {Rousset}, {Roux}, {Salasnich},
  {Sauvage}, {Sevin}, {Soenke}, {Stadler}, {Suarez}, {Weber}, {Wildi},
  {Antoniucci}, {Augereau}, {Baudino}, {Brandner}, {Engler}, {Girard}, {Gry},
  {Kral}, {Kopytova}, {Lagadec}, {Milli}, {Moutou}, {Schlieder},
  {Szul{\'a}gyi}, {Thalmann}, \& {Wahhaj}}]{Chauvin2017}
{Chauvin}, G., {Desidera}, S., {Lagrange}, A.~M., {et~al.} 2017, \aap, 605, L9,
  \dodoi{10.1051/0004-6361/201731152}

\bibitem[{{Crill} {et~al.}(2019){Crill}, {Siegler}, {Bendek}, {Mamajek}, \&
  {Stapelfeldt}}]{Crill2019}
{Crill}, B., {Siegler}, N., {Bendek}, E., {Mamajek}, E., \& {Stapelfeldt}, K.
  2019, in \baas, Vol.~51, 91

\bibitem[{{Currie} {et~al.}(2015){Currie}, {Cloutier}, {Brittain}, {Grady},
  {Burrows}, {Muto}, {Kenyon}, \& {Kuchner}}]{Currie2015}
{Currie}, T., {Cloutier}, R., {Brittain}, S., {et~al.} 2015, \apj, 814, L27,
  \dodoi{10.1088/2041-8205/814/2/L27}

\bibitem[{{Currie} {et~al.}(2014){Currie}, {Daemgen}, {Debes}, {Lafreniere},
  {Itoh}, {Jayawardhana}, {Ratzka}, \& {Correia}}]{Currie2014}
{Currie}, T., {Daemgen}, S., {Debes}, J., {et~al.} 2014, \apjl, 780, L30,
  \dodoi{10.1088/2041-8205/780/2/L30}

\bibitem[{{Currie} {et~al.}(2019{\natexlab{a}}){Currie}, {Pluzhnik}, {Belikov},
  \& {Guyon}}]{Currie2019b}
{Currie}, T., {Pluzhnik}, E., {Belikov}, R., \& {Guyon}, O. 2019{\natexlab{a}},
  in \procspie, Society of Photo-Optical Instrumentation Engineers (SPIE)
  Conference Series.
\newblock \doarXiv{1909.11664}

\bibitem[{{Currie} {et~al.}(2011){Currie}, {Burrows}, {Itoh}, {Matsumura},
  {Fukagawa}, {Apai}, {Madhusudhan}, {Hinz}, {Rodigas}, {Kasper}, {Pyo}, \&
  {Ogino}}]{Currie2011}
{Currie}, T., {Burrows}, A., {Itoh}, Y., {et~al.} 2011, \apj, 729, 128,
  \dodoi{10.1088/0004-637X/729/2/128}

\bibitem[{{Currie} {et~al.}(2012){Currie}, {Debes}, {Rodigas}, {Burrows},
  {Itoh}, {Fukagawa}, {Kenyon}, {Kuchner}, \& {Matsumura}}]{Currie2012}
{Currie}, T., {Debes}, J., {Rodigas}, T.~J., {et~al.} 2012, \apjl, 760, L32,
  \dodoi{10.1088/2041-8205/760/2/L32}

\bibitem[{{Currie} {et~al.}(2019{\natexlab{b}}){Currie}, {Guyon}, {Groff},
  {Kasdin}, {Martinache}, {Brandt}, {Chilcote}, {Marois}, {Jovanovic}, \&
  {Vievard}}]{Currie2019a}
{Currie}, T., {Guyon}, O., {Groff}, T., {et~al.} 2019{\natexlab{b}}, in
  \procspie, Society of Photo-Optical Instrumentation Engineers (SPIE)
  Conference Series.
\newblock \doarXiv{1909.10522}

\bibitem[{{Fitzgerald} {et~al.}(2019){Fitzgerald}, {Bailey}, {Baranec},
  {Batalha}, {Benneke}, {Beichman}, {Brandt}, {Chilcote}, {Chun}, {Crossfield},
  {Currie}, {Davis}, {Dekany}, {Delorme}, {Dong}, {Doyon}, {Dressing},
  {Echeverri}, {Fortney}, {Frazin}, {Guyon}, {Hashimoto}, {Hillenbrand},
  {Hinz}, {Howard}, {Jensen-Clem}, {Jovanovic}, {Kawahara}, {Knutson},
  {Konopacky}, {Kotani}, {Lafreni{\`e}re}, {Liu}, {Lozi}, {Lu}, {Males},
  {Marley}, {Marois}, {Mawet}, {Mazin}, {Millar-Blanchaer}, {Mondal},
  {Murakami}, {Murray-Clay}, {Narita}, {Pezzato}, {Pyo}, {Roberts}, {Ruane},
  {Sallum}, {Serabyn}, {Shields}, {Simard}, {Skemer}, {Stelter}, {Tamura},
  {Troy}, {Vasisht}, {Wallace}, {Wang}, {Wang}, \& {Wright}}]{Fitzgerald2019}
{Fitzgerald}, M., {Bailey}, V., {Baranec}, C., {et~al.} 2019, in \baas,
  Vol.~51, 251

\bibitem[{{Give'on} {et~al.}(2007){Give'on}, {Kern}, {Shaklan}, {Moody}, \&
  {Pueyo}}]{Give'on2007}
{Give'on}, A., {Kern}, B., {Shaklan}, S., {Moody}, D.~C., \& {Pueyo}, L. 2007,
  in Society of Photo-Optical Instrumentation Engineers (SPIE) Conference
  Series, Vol. 6691, \procspie, 66910A, \dodoi{10.1117/12.733122}

\bibitem[{{Greco} \& {Burrows}(2015)}]{GrecoBurrows2015}
{Greco}, J.~P., \& {Burrows}, A. 2015, \apj, 808, 172,
  \dodoi{10.1088/0004-637X/808/2/172}

\bibitem[{{Groff} {et~al.}(2016){Groff}, {Eldorado Riggs}, {Kern}, \& {Jeremy
  Kasdin}}]{Groff2016}
{Groff}, T.~D., {Eldorado Riggs}, A.~J., {Kern}, B., \& {Jeremy Kasdin}, N.
  2016, Journal of Astronomical Telescopes, Instruments, and Systems, 2,
  011009, \dodoi{10.1117/1.JATIS.2.1.011009}

\bibitem[{{Groff} \& {Kasdin}(2013)}]{Groff2013}
{Groff}, T.~D., \& {Kasdin}, N. 2013, Journal of the Optical Society of America
  A, 30, 128, \dodoi{10.1364/JOSAA.30.000128}

\bibitem[{{Guyon}(2003)}]{Guyon2003}
{Guyon}, O. 2003, \aap, 404, 379, \dodoi{10.1051/0004-6361:20030457}

\bibitem[{{Guyon}(2018)}]{Guyon2018c}
---. 2018, \araa, 56, 315, \dodoi{10.1146/annurev-astro-081817-052000}

\bibitem[{{Guyon} {et~al.}(2010){Guyon}, {Martinache}, {Belikov}, \&
  {Soummer}}]{Guyon2010}
{Guyon}, O., {Martinache}, F., {Belikov}, R., \& {Soummer}, R. 2010, \apjs,
  190, 220, \dodoi{10.1088/0067-0049/190/2/220}

\bibitem[{{Guyon} {et~al.}(2018){Guyon}, {Mazin}, {Fitzgerald}, {Mawet},
  {Marois}, {Skemer}, {Lozi}, \& {Males}}]{Guyon2018a}
{Guyon}, O., {Mazin}, B., {Fitzgerald}, M., {et~al.} 2018, in Society of
  Photo-Optical Instrumentation Engineers (SPIE) Conference Series, Vol. 10703,
  Adaptive Optics Systems VI, 107030Z, \dodoi{10.1117/12.2314331}

\bibitem[{{Guyon} {et~al.}(2017){Guyon}, {Miller}, {Males}, {Belikov}, \&
  {Kern}}]{Guyon2017}
{Guyon}, O., {Miller}, K., {Males}, J., {Belikov}, R., \& {Kern}, B. 2017,
  arXiv e-prints, arXiv:1706.07377.
\newblock \doarXiv{1706.07377}

\bibitem[{{Jovanovic} {et~al.}(2015){Jovanovic}, {Martinache}, {Guyon},
  {Clergeon}, {Singh}, {Kudo}, {Garrel}, {Newman}, {Doughty}, {Lozi}, {Males},
  {Minowa}, {Hayano}, {Takato}, {Morino}, {Kuhn}, {Serabyn}, {Norris},
  {Tuthill}, {Schworer}, {Stewart}, {Close}, {Huby}, {Perrin}, {Lacour},
  {Gauchet}, {Vievard}, {Murakami}, {Oshiyama}, {Baba}, {Matsuo}, {Nishikawa},
  {Tamura}, {Lai}, {Marchis}, {Duchene}, {Kotani}, \&
  {Woillez}}]{Jovanovic2015}
{Jovanovic}, N., {Martinache}, F., {Guyon}, O., {et~al.} 2015, \pasp, 127, 890,
  \dodoi{10.1086/682989}

\bibitem[{{Keppler} {et~al.}(2018){Keppler}, {Benisty}, {M{\"u}ller},
  {Henning}, {van Boekel}, {Cantalloube}, {Ginski}, {van Holstein}, {Maire},
  {Pohl}, {Samland }, {Avenhaus}, {Baudino}, {Boccaletti}, {de Boer},
  {Bonnefoy}, {Chauvin}, {Desidera}, {Langlois}, {Lazzoni}, {Marleau},
  {Mordasini}, {Pawellek}, {Stolker}, {Vigan}, {Zurlo}, {Birnstiel},
  {Brandner}, {Feldt}, {Flock}, {Girard}, {Gratton}, {Hagelberg}, {Isella},
  {Janson}, {Juhasz}, {Kemmer}, {Kral}, {Lagrange}, {Launhardt}, {Matter},
  {M{\'e}nard}, {Milli}, {Molli{\`e}re}, {Olofsson}, {P{\'e}rez}, {Pinilla},
  {Pinte}, {Quanz}, {Schmidt}, {Udry}, {Wahhaj}, {Williams}, {Buenzli},
  {Cudel}, {Dominik}, {Galicher}, {Kasper}, {Lannier}, {Mesa}, {Mouillet},
  {Peretti}, {Perrot}, {Salter}, {Sissa}, {Wildi}, {Abe}, {Antichi},
  {Augereau}, {Baruffolo}, {Baudoz}, {Bazzon}, {Beuzit}, {Blanchard}, {Brems},
  {Buey}, {De Caprio}, {Carbillet}, {Carle}, {Cascone}, {Cheetham}, {Claudi},
  {Costille}, {Delboulb{\'e}}, {Dohlen}, {Fantinel}, {Feautrier}, {Fusco},
  {Giro}, {Gluck}, {Gry}, {Hubin}, {Hugot}, {Jaquet}, {Le Mignant}, {Llored},
  {Madec}, {Magnard}, {Martinez}, {Maurel}, {Meyer}, {M{\"o}ller-Nilsson},
  {Moulin}, {Mugnier}, {Orign{\'e}}, {Pavlov}, {Perret}, {Petit}, {Pragt},
  {Puget}, {Rabou}, {Ramos}, {Rigal}, {Rochat}, {Roelfsema}, {Rousset}, {Roux},
  {Salasnich}, {Sauvage}, {Sevin}, {Soenke}, {Stadler}, {Suarez}, {Turatto}, \&
  {Weber}}]{Keppler2018}
{Keppler}, M., {Benisty}, M., {M{\"u}ller}, A., {et~al.} 2018, \aap, 617, A44,
  \dodoi{10.1051/0004-6361/201832957}

\bibitem[{{Lagrange} {et~al.}(2010){Lagrange}, {Bonnefoy}, {Chauvin}, {Apai},
  {Ehrenreich}, {Boccaletti}, {Gratadour}, {Rouan}, {Mouillet}, {Lacour}, \&
  {Kasper}}]{Lagrange2010}
{Lagrange}, A.-M., {Bonnefoy}, M., {Chauvin}, G., {et~al.} 2010, Science, 329,
  57, \dodoi{10.1126/science.1187187}

\bibitem[{{Lopez-Morales} {et~al.}(2019){Lopez-Morales}, {Currie}, {Teske},
  {Gaidos}, {Kempton}, {Males}, {Lewis}, {Rackham}, {Ben-Ami}, {Birkby},
  {Charbonneau}, {Close}, {Crane}, {Dressing}, {Froning}, {Hasegawa},
  {Konopacky}, {Kopparapu}, {Mawet}, {Mennesson}, {Ramirez}, {Stelter},
  {Szentgyorgyi}, {Wang}, {Alam}, {Collins}, {Dupree}, {Karovska}, {Kirk},
  {Levi}, {McGruder}, {Packman}, {Rugheimer}, \& {Rukdee}}]{LopezMorales2019}
{Lopez-Morales}, M., {Currie}, T., {Teske}, J., {et~al.} 2019, in \baas,
  Vol.~51, Bulletin of the American Astronomical Society, 162.
\newblock \doarXiv{1903.09523}

\bibitem[{{Lozi} {et~al.}(2018){Lozi}, {Guyon}, {Jovanovic}, {Goebel},
  {Pathak}, {Skaf}, {Sahoo}, {Norris}, {Martinache}, {N'Diaye}, {Mazin},
  {Walter}, {Tuthill}, {Kudo}, {Kawahara}, {Kotani}, {Ireland}, {Cvetojevic},
  {Huby}, {Lacour}, {Vievard}, {Groff}, {Chilcote}, {Kasdin}, {Knight}, {Snik},
  {Doelman}, {Minowa}, {Clergeon}, {Takato}, {Tamura}, {Currie}, {Takami}, \&
  {Hayashi}}]{Lozi2018}
{Lozi}, J., {Guyon}, O., {Jovanovic}, N., {et~al.} 2018, in Society of
  Photo-Optical Instrumentation Engineers (SPIE) Conference Series, Vol. 10703,
  \procspie, 1070359, \dodoi{10.1117/12.2314282}

\bibitem[{{Macintosh} {et~al.}(2015){Macintosh}, {Graham}, {Barman}, {De Rosa},
  {Konopacky}, {Marley}, {Marois}, {Nielsen}, {Pueyo}, {Rajan}, {Rameau},
  {Saumon}, {Wang}, {Patience}, {Ammons}, {Arriaga}, {Artigau}, {Beckwith},
  {Brewster}, {Bruzzone}, {Bulger}, {Burningham}, {Burrows}, {Chen}, {Chiang},
  {Chilcote}, {Dawson}, {Dong}, {Doyon}, {Draper}, {Duch{\^e}ne}, {Esposito},
  {Fabrycky}, {Fitzgerald}, {Follette}, {Fortney}, {Gerard}, {Goodsell},
  {Greenbaum}, {Hibon}, {Hinkley}, {Cotten}, {Hung}, {Ingraham},
  {Johnson-Groh}, {Kalas}, {Lafreniere}, {Larkin}, {Lee}, {Line}, {Long},
  {Maire}, {Marchis}, {Matthews}, {Max}, {Metchev}, {Millar-Blanchaer},
  {Mittal}, {Morley}, {Morzinski}, {Murray-Clay}, {Oppenheimer}, {Palmer},
  {Patel}, {Perrin}, {Poyneer}, {Rafikov}, {Rantakyr{\"o}}, {Rice}, {Rojo},
  {Rudy}, {Ruffio}, {Ruiz}, {Sadakuni}, {Saddlemyer}, {Salama}, {Savransky},
  {Schneider}, {Sivaramakrishnan}, {Song}, {Soummer}, {Thomas}, {Vasisht},
  {Wallace}, {Ward- Duong}, {Wiktorowicz}, {Wolff}, \&
  {Zuckerman}}]{Macintosh2015}
{Macintosh}, B., {Graham}, J.~R., {Barman}, T., {et~al.} 2015, Science, 350,
  64, \dodoi{10.1126/science.aac5891}

\bibitem[{{Malbet} {et~al.}(1995){Malbet}, {Yu}, \& {Shao}}]{Malbet1995}
{Malbet}, F., {Yu}, J.~W., \& {Shao}, M. 1995, \pasp, 107, 386,
  \dodoi{10.1086/133563}

\bibitem[{{Males} \& {Guyon}(2018)}]{MalesGuyon2018}
{Males}, J.~R., \& {Guyon}, O. 2018, Journal of Astronomical Telescopes,
  Instruments, and Systems, 4, 019001, \dodoi{10.1117/1.JATIS.4.1.019001}

\bibitem[{{Males} {et~al.}(2018){Males}, {Close}, {Miller}, {Schatz},
  {Doelman}, {Lumbres}, {Snik}, {Rodack}, {Knight}, {Van Gorkom}, {Long},
  {Hedglen}, {Kautz}, {Jovanovic}, {Morzinski}, {Guyon}, {Douglas}, {Follette},
  {Lozi}, {Bohlman}, {Durney}, {Gasho}, {Hinz}, {Ireland}, {Jean}, {Keller},
  {Kenworthy}, {Mazin}, {Noenickx}, {Alfred}, {Perez}, {Sanchez}, {Sauve},
  {Weinberger}, \& {Conrad}}]{Males2018}
{Males}, J.~R., {Close}, L.~M., {Miller}, K., {et~al.} 2018, in Society of
  Photo-Optical Instrumentation Engineers (SPIE) Conference Series, Vol. 10703,
  \procspie, 1070309, \dodoi{10.1117/12.2312992}

\bibitem[{{Marois} {et~al.}(2008){Marois}, {Macintosh}, {Barman}, {Zuckerman},
  {Song}, {Patience}, {Lafreni{\`e}re}, \& {Doyon}}]{Marois2008}
{Marois}, C., {Macintosh}, B., {Barman}, T., {et~al.} 2008, Science, 322, 1348,
  \dodoi{10.1126/science.1166585}

\bibitem[{{Mawet} {et~al.}(2014){Mawet}, {Milli}, {Wahhaj}, {Pelat}, {Absil},
  {Delacroix}, {Boccaletti}, {Kasper}, {Kenworthy}, {Marois}, {Mennesson}, \&
  {Pueyo}}]{Mawet2014}
{Mawet}, D., {Milli}, J., {Wahhaj}, Z., {et~al.} 2014, \apj, 792, 97,
  \dodoi{10.1088/0004-637X/792/2/97}

\bibitem[{{Miller} {et~al.}(2017){Miller}, {Guyon}, \& {Males}}]{Miller2017}
{Miller}, K., {Guyon}, O., \& {Males}, J. 2017, Journal of Astronomical
  Telescopes, Instruments, and Systems, 3, 049002,
  \dodoi{10.1117/1.JATIS.3.4.049002}

\bibitem[{{Miller} {et~al.}(2019){Miller}, {Males}, {Guyon}, {Close},
  {Doelman}, {Snik}, {Por}, {Wilby}, {Keller}, {Bohlman}, {Van Gorkom},
  {Rodack}, {Knight}, {Lumbres}, {Bos}, \& {Jovanovic}}]{Miller2019}
{Miller}, K., {Males}, J.~R., {Guyon}, O., {et~al.} 2019, Journal of
  Astronomical Telescopes, Instruments, and Systems, 5, 049004,
  \dodoi{10.1117/1.JATIS.5.4.049004}

\bibitem[{{N'Diaye} {et~al.}(2013){N'Diaye}, {Dohlen}, {Fusco}, \&
  {Paul}}]{Ndiaye2013}
{N'Diaye}, M., {Dohlen}, K., {Fusco}, T., \& {Paul}, B. 2013, \aap, 555, A94,
  \dodoi{10.1051/0004-6361/201219797}

\bibitem[{{N'Diaye} {et~al.}(2016){N'Diaye}, {Vigan}, {Dohlen}, {Sauvage},
  {Caillat}, {Costille}, {Girard}, {Beuzit}, {Fusco}, {Blanchard}, {Le Merrer},
  {Le Mignant}, {Madec}, {Moreaux}, {Mouillet}, {Puget}, \&
  {Zins}}]{Ndiaye2016}
{N'Diaye}, M., {Vigan}, A., {Dohlen}, K., {et~al.} 2016, \aap, 592, A79,
  \dodoi{10.1051/0004-6361/201628624}

\bibitem[{{Pluzhnik} {et~al.}(2017){Pluzhnik}, {Sirbu}, {Belikov}, {Bendek}, \&
  {Dudinov}}]{Pluzhnik2017}
{Pluzhnik}, E., {Sirbu}, D., {Belikov}, R., {Bendek}, E., \& {Dudinov}, V.~N.
  2017, arXiv e-prints, arXiv:1709.01571.
\newblock \doarXiv{1709.01571}

\bibitem[{{Pueyo} \& {Kasdin}(2007)}]{Pueyo2007}
{Pueyo}, L., \& {Kasdin}, N.~J. 2007, \apj, 666, 609, \dodoi{10.1086/518884}

\bibitem[{{Rajan} {et~al.}(2017){Rajan}, {Rameau}, {De Rosa}, {Marley},
  {Graham}, {Macintosh}, {Marois}, {Morley}, {Patience}, {Pueyo}, {Saumon},
  {Ward-Duong}, {Ammons}, {Arriaga}, {Bailey}, {Barman}, {Bulger}, {Burrows},
  {Chilcote}, {Cotten}, {Czekala}, {Doyon}, {Duch{\^e}ne}, {Esposito},
  {Fitzgerald}, {Follette}, {Fortney}, {Goodsell}, {Greenbaum}, {Hibon},
  {Hung}, {Ingraham}, {Johnson-Groh}, {Kalas}, {Konopacky}, {Lafreni{\`e}re},
  {Larkin}, {Maire}, {Marchis}, {Metchev}, {Millar-Blanchaer}, {Morzinski},
  {Nielsen}, {Oppenheimer}, {Palmer}, {Patel}, {Perrin}, {Poyneer},
  {Rantakyr{\"o}}, {Ruffio}, {Savransky}, {Schneider}, {Sivaramakrishnan},
  {Song}, {Soummer}, {Thomas}, {Vasisht}, {Wallace}, {Wang}, {Wiktorowicz}, \&
  {Wolff}}]{Rajan2017}
{Rajan}, A., {Rameau}, J., {De Rosa}, R.~J., {et~al.} 2017, \aj, 154, 10,
  \dodoi{10.3847/1538-3881/aa74db}

\bibitem[{{Rameau} {et~al.}(2013){Rameau}, {Chauvin}, {Lagrange}, {Meshkat},
  {Boccaletti}, {Quanz}, {Currie}, {Mawet}, {Girard}, {Bonnefoy}, \&
  {Kenworthy}}]{Rameau2013}
{Rameau}, J., {Chauvin}, G., {Lagrange}, A.~M., {et~al.} 2013, \apjl, 779, L26,
  \dodoi{10.1088/2041-8205/779/2/L26}

\bibitem[{{Seo} {et~al.}(2018){Seo}, {Shi}, {Balasubramanian}, {Cady},
  {Gordon}, {Kern}, {Lam}, {Marx}, {Moody}, {Muller}, {Patterson},
  {Poberezhskiy}, {Mejia Prada}, {Riggs}, {Trauger}, \& {Wilson}}]{Seo2018}
{Seo}, B.-J., {Shi}, F., {Balasubramanian}, B., {et~al.} 2018, in Society of
  Photo-Optical Instrumentation Engineers (SPIE) Conference Series, Vol. 10698,
  \procspie, 106982P, \dodoi{10.1117/12.2314358}

\bibitem[{{Shaklan} \& {Green}(2006)}]{ShaklanGreen2006}
{Shaklan}, S.~B., \& {Green}, J.~J. 2006, \ao, 45, 5143,
  \dodoi{10.1364/AO.45.005143}

\bibitem[{{Shi} {et~al.}(2018){Shi}, {Seo}, {Cady}, {Kern}, {Lam}, {Marx},
  {Patterson}, {Mejia Prada}, {Shaw}, {Shelton}, {Shields}, {Tang}, \&
  {Truong}}]{Shi2018}
{Shi}, F., {Seo}, B.-J., {Cady}, E., {et~al.} 2018, in Society of Photo-Optical
  Instrumentation Engineers (SPIE) Conference Series, Vol. 10698, \procspie,
  106982O, \dodoi{10.1117/12.2312746}

\bibitem[{{Soummer} {et~al.}(2011){Soummer}, {Hagan}, {Pueyo}, {Thormann},
  {Rajan}, \& {Marois}}]{Soummer2011}
{Soummer}, R., {Hagan}, J.~B., {Pueyo}, L., {et~al.} 2011, \apj, 741, 55,
  \dodoi{10.1088/0004-637X/741/1/55}

\bibitem[{{Sun} {et~al.}(2020){Sun}, {Kasdin}, \& {Vanderbei}}]{Sun2020}
{Sun}, H., {Kasdin}, N.~J., \& {Vanderbei}, R. 2020, Journal of Astronomical
  Telescopes, Instruments, and Systems, 6, 019001,
  \dodoi{10.1117/1.JATIS.6.1.019001}

\bibitem[{{Trauger} {et~al.}(2011){Trauger}, {Moody}, {Gordon}, {Krist}, \&
  {Mawet}}]{Trauger2011}
{Trauger}, J., {Moody}, D., {Gordon}, B., {Krist}, J., \& {Mawet}, D. 2011,
  Society of Photo-Optical Instrumentation Engineers (SPIE) Conference Series,
  Vol. 8151, {A hybrid Lyot coronagraph for the direct imaging and spectroscopy
  of exoplanet systems: recent results and prospects}, 81510G,
  \dodoi{10.1117/12.895032}

\end{thebibliography}
\bibliographystyle{aasjournal} % makes bibtex use spiebib.bst

\end{document}